\documentclass[aps,prd,twocolumn,superscriptaddress,nofootinbib]{revtex4-1}

\usepackage{latexsym}
\usepackage{amsmath}
\usepackage{amssymb}
\usepackage{amsfonts}
\usepackage{bm}

\usepackage{color}

\usepackage{supertabular} 
\usepackage{placeins}
\usepackage{epsfig}
\usepackage{graphicx}

\begin{document}

%\preprint{}

\title{Doubly-heavy tetraquarks}


\author{Gang Yang }
\email[]{ygz0788a@sina.com}
%\thanks{}
\affiliation{Department of Physics and State Key Laboratory of Low-Dimensional Quantum Physics, \\ Tsinghua University, Beijing 100084, P. R. China}

\author{Jialun Ping}
\email[]{jlping@njnu.edu.cn}
\affiliation{Department of Physics and Jiangsu Key Laboratory for Numerical Simulation of Large Scale Complex Systems, Nanjing Normal University, Nanjing 210023, P. R. China}

%\author{Jorge Segovia}
%\email[]{jsegovia@ifae.es}
%\affiliation{Institut de F\'isica d'Altes Energies (IFAE) and Barcelona Institute of Science and Technology (BIST), \\ Universitat Aut\`onoma de Barcelona, E-08193 Bellaterra (Barcelona), Spain}
%
\author{Jorge Segovia}
\email[]{jsegovia@upo.es}
\affiliation{Departamento de Sistemas F\'isicos, Qu\'imicos y Naturales, \\ Universidad Pablo de Olavide, E-41013 Sevilla, Spain}


%%%%%%%%%\date{\today}

\begin{abstract}
In the framework of the chiral quark model along with complex scaling range, we perform a dynamical  study on the low-lying $S$-wave doubly-heavy tetraquark states ($QQ\bar{q}\bar{q}$, $Q=c, b$ and $q=u, d$) with an accurate computing approach, Gaussian expansion method. The meson-meson and diquark-antidiquark configurations within all possible color structures for spin-parity quantum numbers $J^P=0^+$, $1^+$ and $2^+$, and in the $0$ and $1$ isospin sectors are considered. 
Possible tightly bound and narrow resonance states are obtained for doubly-charm and doubly-bottom tetraquarks with $IJ^P=01^+$, and these exotic states are also obtained in charm-bottom tetraquarks with $00^+$ and $01^+$ quantum numbers. Only loosely bound state is found in charm-bottom tetraquarks of $02^+$ states. All of these bound states within meson-meson configurations are loosely bound whether in color-singlet channels or coupling to hidden-color ones. However compact structures are available in diquark-antidiquark channels except for charm-bottom tetraquarks in $02^+$ states.
\end{abstract}


\maketitle

%%%%%%%%%%%%%%%%%%%%%%%%%%%%%%%%%%%%%%%%%%%%%%%%%%%%%%%%%%%%%%%%%%%%%%%%%%%%%%%%%%%%%%%%%%

\section{Introduction}

The story of exotic hadronic states can be dated back to the announcement of $X(3872)$ in the invariant mass spectrum of $J/\psi \pi^+ \pi^-$ produced in $B^\pm \rightarrow K^\pm X(3872) \rightarrow K^\pm J/\psi \pi^+ \pi^-$ decays by the Belle Collaboration in 2003~\cite{skc:2003prl}. This charmonium-like state was confirmed by other experimental collaborations~\cite{da:2004prl, vma:2004prl, ba:2005prd} during the following years. However, theoretical explanations on $X(3872)$ are still controversial: (i) in spit of the predicted mass of $\chi_{c1}(2P)$ is too high ($\sim3.95\,\text{GeV}$)  to identified with $X(3872)$~\cite{sg:1985prd, de:2011epjc, Segovia:2013wma, Vijande:2004he}, the radiative decays are better described in charmonium structure~\cite{tb:2004prd, eje:2004prd}, (ii) mass near the $D^0\bar{D}^{*0}$ threshold is compatible with molecular state~\cite{ybd:2008prd, ybd:2009prd, dg:2010prd, fkg:2015plb} and the comprehensible isospin breaking decay process of $X(3872)\rightarrow J/\psi \rho$, (iii) the $X(3872)$ is also described as a compact diquark-antidiquark state~\cite{lm:2005prd}, and (iv) the existence of $c\bar{c}$ bound states dressed by $DD^*$ moleucular component is proposed~\cite{pgo:2010prd, scgr:2011epjc, jfgg:2014prd, mcgr:2011epjc, ytjp:2019prd}. In fact, during the past 16 years, more than two dozens of unconventional charmonium- and bottomonium-like states, the so-called XYZ mesons, have been observed at B-factories (BaBar, Belle and CLEO), $\tau$-charm facilities (CLEO-c and BESIII) and also proton-(anti)proton colliders (CDF, D0, LHCb, ATLAS and CMS), $e. g.$ $Y(4260)$ discovered by the BaBar Collaboration in 2005~\cite{ba:2005prl}, $Z^+(4430)$ discovered by the Belle Collaboration in 2007~\cite{skcslo:2008prl}, $Y(4140)$ discovered by the CDF Collaboration in 2009~\cite{ta:2009prl}, and $Z^+_c(3900)$ discovered by the BESIII Collaboration in 2013~\cite{ta:2013prl}, $etc.$ Meanwhile, remarkable achievements in the baryon sectors are also valuable. In 2015 two exotic hidden-charmonium pentaquarks, $P^+_c(4380)$ and $P^+_c(4450)$ were announced by the LHCb Collaboration~\cite{Aaij:2015tga} in the $\Lambda^{0}_{b}$ decay, $\Lambda^{0}_{b} \rightarrow J/\psi K^{-}p$ and in 2019 with higher statistical significance, one new pentaquark state $P^+_c(4312)$ was found by the same collaboration and the previously reported wide state $P^+_c(4450)$ was superseded by two narrow ones, $P^+_c(4440)$ and $P^+_c(4457)$~\cite{lhcb:2019pc}. Review on these exotic states can be found in Ref.~\cite{hxc:2016pr, hxc:2017rpp, fkg:2018rmp, slo:2018rmp}.

Apparently, these facts have triggered large amount of theoretical investigations on the new hadronic zoo where the conventional configuration of mesons and baryons as, respectively, quark-antiquark and $3$-quark bound states is being left behind. In fully-heavy tetraquarks sector, the CMS collaboration claimed an observation of pair production of $\Upsilon(1S)$ mesons at the LHC in $pp$ collisions~\cite{vkams:2017jhep} and this may indicated a $bb\bar{b}\bar{b}$ tetraquark state with mass of $18.4\,\text{GeV}$. A significant peak at $\sim$$18.2\,\text{GeV}$ was observed in Cu+Au collisions at RHIC~\cite{LCBland2019}. However, no evidence has been provided from the LHCb collaboration by searching for the $\Upsilon(1S)\mu^+ \mu^-$ invariant mass spectrum~\cite{raba:2018jhep}. Extensive theoretical works with different schemes devote to these extremely non-relativistic systems, $QQ\bar{Q}\bar{Q}$ $(Q=c, b)$: the existence of $bb\bar{b}\bar{b}$ bound state is supported by phenomenological model calculation~\cite{avb:2012prd, mna:2018epjc, aeap:2018epjc, mabjfcdres2019}, QCD sum rules~\cite{zgwqqqq:2017epjc, wchxc:2017plb}, and diffusion Monte Carlo method~\cite{yb:2019plb}. A narrow $cc\bar{c}\bar{c}$ tetraquark state in the mass region $5-\,6\,\text{GeV}$ has been predicted by the Bethe-Salpeter approach~\cite{whge:2012plb} and also in several phenomenological models~\cite{avb:2012prd, vrdfsn:2019cpc, avbakl:2011prd, mksnjl:2017prd}. However, there are still intense debates on the observation of these exotic states. No $cc\bar{c}\bar{c}$ and $bb\bar{b}\bar{b}$ bound states can be formed within effective model calculations~\cite{jmrav:2017prd, jwyrl:2018prd, xc:2019epja, mslqfl:2019prd, gjw:2019arx, jmravjv2018} and lattice QCD~\cite{cheec:2018prd}, but possible stable or narrow states in the $bb\bar{b}\bar{c}$ and $bc\bar{b}\bar{c}$ systems~\cite{jmrav:2017prd, jwyrl:2018prd}.

Nevertheless, results on doubly-heavy tetraquark states investigated by different kinds of theoretical approaches are more compatible. In heavy quark limit, stable and extremely narrow $bb\bar{u}\bar{d}$ tetraquark state with the $J^P=1^+$ must exist~\cite{ejecq:2017prl}. In Ref.~\cite{mkjlr:2017prl} the predicted mass of  $bb\bar{u}\bar{d}$ state within the same spin-parity is $10389\pm 12\,\text{MeV}$. Mass, lifetime and decay modes of this tetraquark are investigated in Ref.~\cite{EHJVAVJMR2019}. A compact doubly-bottom tetraquark state with $IJ^P=01^+$ is also presented in heavy-ion collisions at the LHC~\cite{cefgk:2019prd} and actually, in 1988 the dimeson $T(bb\bar{u}\bar{d})$ had already been proposed~\cite{jclh:1988prd}. Besides, a narrow $(bb)(\bar{u}\bar{d})$ diquark-antidiquark state with $IJ^P=01^+$ is   predicted in Ref.~\cite{dernfvogwl2007}. A $\bar{b}\bar{b}ud$ bound state also with $IJ^P=01^+$ is stable against the  strong and electromagnetic decay and its mass is $10476\pm 24\pm 10\,\text{MeV}$ by Lattice QCD~\cite{llsm:2019prd}, this deeply bound state is supported also by the same formalism in Refs.~\cite{afrjhrlkm2017, pjnmmp2019}. Meanwhile, there are also QCD sum rules predicted a mass $7105\pm 155\,\text{MeV}$ for $bc\bar{u}\bar{d}$ axial-vector tetraquark state~\cite{ssaka:2019arx}, and $I(J^P)=0(1^+)$ $ud\bar{c}\bar{d}$ tetraquark which binding energy is 15 to $61\,\text{MeV}$ with respect to $\bar{D}B^*$ threshold is proposed by Ref.~\cite{afrjhrlkm2019}. Moreover, the production potential of doubly-heavy tetraquarks at a Tera-Z factory and the LHC are estimated by Monte Carlo simulation~\cite{aaaypqqww2018, aaqqww2018}.
 However, no strong indication for any bound state or narrow resonance of tetraquarks in charm sector are found in Lattice study~\cite{gkcccetjjdrge2017}. 
 Some other types of tetraquark states along with decay properties are explored in Refs.~\cite{ssakahs:2019prd, yyjp:2019prd, zgw:2019arx}.

We study herein, within a complex scaling range of chiral quark model formalism, the possibility of having tetraquark bound- and resonance-states in the doubly-heavy sector with quantum numbers $J^P=0^+$, $1^+$ and $2^+$, and in the $0$ and $1$ isospin sectors. Two configurations, meson-meson and diquark-antidiquark structures are considered. In particular, color-singlet and hidden-color channels for dimeson configuration, color triplet-antitriple and sextet-antisextet channels for diquark-antidiquark one along with their couplings are all employed for each quantum states. The bound states, if possible, their internal structures and components in the complete coupled-channels calculation are analyzed by computing the distances among any pair of quarks and the contributions of each channel's wave functions. Meanwhile, masses and widths for possible resonance states are also studied in the complete coupled-channels. 

%All the details about our computational framework will be described later but let us sketch here some of its main features. Our chiral quark model (ChQM) is based on the fact that chiral symmetry is spontaneously broken in QCD and, among other consequences, it provides a constituent quark mass to the light quarks. To restore the chiral symmetry in the QCD Lagrangian, Goldstone-boson exchange interactions appear between the light quarks. This fact is encoded in a phenomenological potential which already contains the perturbative one-gluon exchange (OGE) interaction and a nonperturbative linear-screened confining term.\footnote{The interested reader is referred to Refs.~\cite{Valcarce:2005em, Segovia:2013wma} for detailed reviews on the naive quark model in which this work is based.} It is worth to note that chiral symmetry is explicitly broken in the  heavy quark sector and this translates in our formalism to the fact that the interaction terms between light-light, light-heavy and heavy-heavy quarks are not the same, i.e. while Goldstone-boson exchanges are considered when the two quarks are light, they do not appear in the other two configurations: light-heavy and heavy-heavy; however, the one-gluon exchange and confining potentials are flavor blindness.

\begin{figure}[ht]
\epsfxsize=3.4in \epsfbox{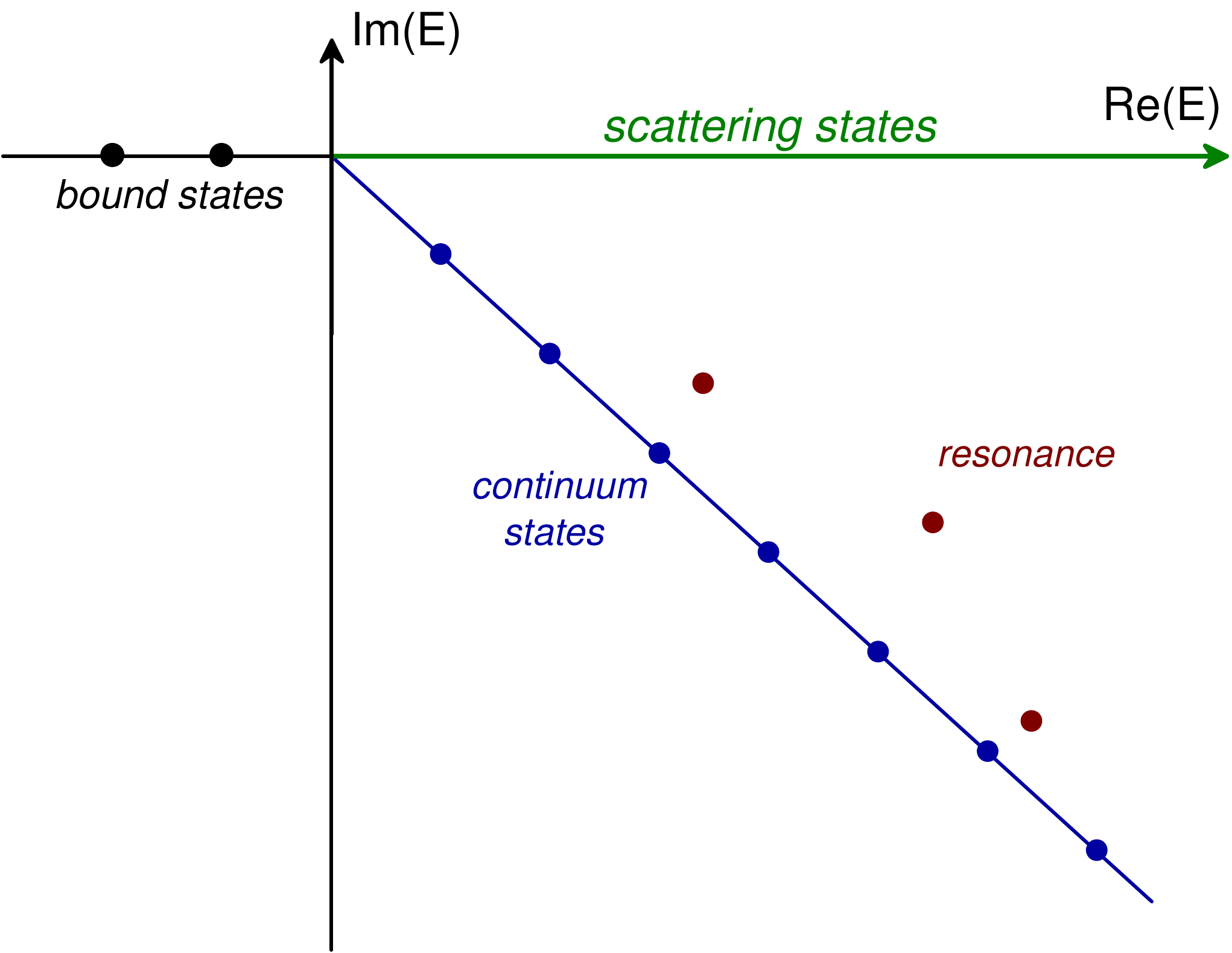}
\caption{Schematic complex energy distribution in the single-channel two-body system.} \label{CSM1}
\end{figure}

The four-body bound state problem is implemented by two strong foundations, the Gaussian expansion method (GEM)~\cite{Hiyama:2003cu} which has been demonstrated to be as accurate as a Faddeev calculation (see, for instance, Figs.~15 and~16 of Ref.~\cite{Hiyama:2003cu}), and the chiral quark model which has been successfully applied to hadron~\cite{Valcarce:1995dm, Vijande:2004he, Segovia:2008zza, Segovia:2008zz, Ortega:2016hde, Yang:2017xpp, gy:2019cpc}, hadron-hadron ~\cite{Fernandez:1993hx, Valcarce:1994nr, Ortega:2009hj, Ortega:2016mms, Ortega:2016pgg} and multiquark~\cite{Vijande:2006jf, Yang:2015bmv, Yang:2017rpg, gy:2019prdnn} phenomenology.
However, due to the complexity of the coupled channels case for scattering and resonance states, it is difficult to solve a scattering issue together with resonance one. In this work, a powerful technique, complex scaling method (CSM) is employed, and this is also for the first time that its application to tetraquark states in hadronic physics. During the past decades, it has been extensively applied to nuclear physics problems~\cite{SAPTP11612006, TMPPNP7912014}, and recently also in the study of charmed dibaryon resonances~\cite{MOSMYRL2019}. The CSM is quite different from a real range one, for the scattering, resonance and bound states can all be concordant in one calculation (see Fig.~\ref{CSM1}, a schematic distribution of the complex energy of 2-body by the CSM according to Ref.~\cite{TMPPNP7912014} ), namely the scattering states can be solved as a bound states problem without Lippmann-Schwinger equation or some scattering issues related, and the resonance pole will be fixed in the complex plane. A briefly sketch for the application of CSM in tetraquark states will be shown in the next section.

The structure of this paper is organized in the following way. In Sec.~\ref{sec:model} theoretical framework which includes the ChQM, tetraquark wave-functions, GEM and CSM is briefly presented and discussed. Section~\ref{sec:results} is devoted to the analysis and discussion on the obtained results. The summary and some prospects are presented in Sec.~\ref{sec:summary}.

%%%%%%%%%%%%%%%%%%%%%%%%%%%%%%%%%%%%%%%%%%%%%%%%%%%%%%%%%%%%%%%%%%%%%%%%%%%%%%%%%%%%%%%%%%

\section{Theoretical framework}
\label{sec:model}

With half a century development in high energy physics, the QCD-inspired quark models are still the main tool to shed some light on the nature of the multiquark candidates observed by experimentalists. Particularly, the chiral quark model has been witnessed great achievements in our early work on possible hidden-charm pentaquark bound states with quantum numbers $IJ^P=\frac{1}{2}\left(\frac{1}{2}\right)^\pm$, $\frac{1}{2}\left(\frac{3}{2}\right)^\pm$ and $\frac{1}{2}\left(\frac{5}{2}\right)^\pm$~\cite{Yang:2015bmv}. Therein, the properties were compared with those associated with the hidden-charm pentaquark signals observed by the LHCb Collaboration in 2015~\cite{Aaij:2015tga}. 
%Furthermore, one can find more inspiring results that before the announcement of three new hidden-charm pentaquarks  which were also discovered by the same collaboration~\cite{lhcb:2019pc}, their masses had already been nicely predicted in Ref.~\cite{Yang:2015bmv} (see Tables III and IV of Ref.~\cite{Yang:2015bmv}), $P^+_c(4312)$, $P^+_c(4440)$ and $P^+_c(4457)$ can be explained as the molecular states of $J^P=\frac12^-$ $\Sigma_c\bar{D}$, $\frac12^-$ $\Sigma_c\bar{D}^*$ and $\frac32^-$ $\Sigma_c\bar{D}^*$, respectively and the isospin are all of $\frac12$; moreover our results are supported by Refs.~\cite{MZL190311560, JH190311872, CWX190401296, CJX190400872}.
Although three new hidden-charm pentaquarks were also reported by the same collaboration in 2019~\cite{lhcb:2019pc}, these states are not discussing exactly in the present work. Herein, the application of chiral quark model in doubly-heavy tetraquark states is quite expected.

%Although Lattice QCD (LQCD) has made an impressive progress on understanding multiquark systems~\cite{Alexandrou:2001ip, Okiharu:2004wy} and the hadron-hadron interaction~\cite{Prelovsek:2014swa, Lang:2014yfa, Briceno:2017max}, the QCD-inspired quark models are still the main tool to shed some light on the nature of the multiquark candidates observed by experimentalists.

The general form of our four-body Hamiltonian in complex scaling method is
\begin{equation}
H(\theta) = \sum_{i=1}^{4}\left( m_i+\frac{\vec{p\,}^2_i}{2m_i}\right) - T_{\text{CM}} + \sum_{j>i=1}^{4} V(\vec{r}_{ij} e^{i\theta}) \,,
\label{eq:Hamiltonian}
\end{equation}
where the center-of-mass kinetic energy $T_{\text{CM}}$ is subtracted without losing a generality
since we mainly focus on the internal relative motions of multi-quark system. Interaction part is of two-body potential
\begin{equation}\label{CQMV}
V(\vec{r}_{ij} e^{i\theta}) = V_{\text{CON}}(\vec{r}_{ij} e^{i\theta}) + V_{\text{OGE}}(\vec{r}_{ij} e^{i\theta}) +
    V_{\chi}(\vec{r}_{ij} e^{i\theta}) \,,
\end{equation}
includes the color-confining, one-gluon exchange and Goldstone-boson exchange interactions. Note herein that only the central and spin-spin of potential are considered since our main goal of the present work is to perform a systematical study on the low-lying $S$-wave doubly-heavy tetraquark states, it is reasonable for the absence of spin-orbit and tensor contributions.
One can see that the coordinates of relative motions between quarks are transformed with a complex rotation, $\vec{r} \rightarrow \vec{r} e^{i\theta}$. Accordingly, in the framework of complex range, the four-body systems are solved in a complex scaled Schr\"{o}dinger euqation:
\begin{equation}\label{CSMSE}
\left[ H(\theta)-E(\theta) \right] \Psi_{JM}(\theta)=0	
\end{equation}

According to the ABC theorem~\cite{JA22269, EB22280}, there are three types of complex eigenenergies of Eq.~(\ref{CSMSE}) as shown in Fig.~\ref{CSM1}:

(1) The bound state below the threshold is always located on the negative axis of real energy.

(2) The discretized continuum state are aligned along the cut line with a rotated angle of 2$\theta$ related to the real axis.

(3) The resonance state is a fixed pole under the complex scaling transformation and is located above the continuum cut line. The resonance width is given by $\Gamma=-2Im(E)$.

As an illustration to each interaction potentials in Eq.~(\ref{CQMV}). Firstly, color confinement should be encoded in the non-Abelian character of QCD. It has been demonstrated by LQCD that multi-gluon exchanges produce an attractive linearly rising potential proportional to the distance between infinite-heavy quarks~\cite{Bali:2005fu}. However, the spontaneous creation of light-quark pairs from the QCD vacuum may give rise at the same scale to a breakup of the created color flux-tube~\cite{Bali:2005fu}. These two phenomenological observations are mimicked by the following expression when $\theta=0^\circ$:
\begin{equation}
V_{\text{CON}}(\vec{r}_{ij} e^{i\theta}\,)=\left[-a_{c}(1-e^{-\mu_{c}r_{ij} e^{i\theta}})+\Delta \right] 
(\vec{\lambda}_{i}^{c}\cdot\vec{\lambda}_{j}^{c}) \,,
\label{eq:conf}
\end{equation}
where $a_{c}$, $\mu_{c}$ and $\Delta$ are model parameters, and the SU(3) color Gell-Mann matrices are denoted as $\lambda^c$. One can see in Eq.~\eqref{eq:conf} that the potential is linear at short inter-quark distances with an effective confinement strength $\sigma = -a_{c} \, \mu_{c} \, (\vec{\lambda}^{c}_{i}\cdot \vec{\lambda}^{c}_{j})$, while $V_{\text{CON}}$ becomes constant $(\Delta-a_c)(\vec{\lambda}^{c}_{i}\cdot \vec{\lambda}^{c}_{j})$ at large distances. 

The one-gluon exchange potential which includes the coulomb and color-magnetism interactions is given by
\begin{align}
&
V_{\text{OGE}}(\vec{r}_{ij} e^{i\theta}) = \frac{1}{4} \alpha_{s} (\vec{\lambda}_{i}^{c}\cdot
\vec{\lambda}_{j}^{c}) \Bigg[\frac{1}{r_{ij} e^{i\theta}} \nonumber \\ 
&
\hspace*{1.60cm} - \frac{1}{6m_{i}m_{j}} (\vec{\sigma}_{i}\cdot\vec{\sigma}_{j}) 
\frac{e^{-r_{ij} e^{i\theta}/r_{0}(\mu)}}{r_{ij} e^{i\theta} r_{0}^{2}(\mu)} \Bigg] \,,
\end{align}
where $m_{i}$ and $\vec{\sigma}$ are the quark mass and the Pauli matrices respectively. The contact term of the central potential in complex range has been regularized as
\begin{equation}
\delta(\vec{r}_{ij} e^{i\theta})\sim\frac{1}{4\pi r_{0}^{2}}\frac{e^{-r_{ij} e^{i\theta}/r_{0}}}{r_{ij} e^{i\theta}} \,,
\end{equation}
with $r_{0}(\mu_{ij})=\hat{r}_{0}/\mu_{ij}$ a regulator that depends on $\mu_{ij}$, the reduced mass of the quark--(anti-)quark pair.

The QCD strong coupling constant $\alpha_s$ (an effective scale-dependent strong coupling constant) offers a consistent description of mesons and baryons from light to heavy quark sectors in wide energy range, and we use the frozen coupling constant of, for instance, Ref.~\cite{Segovia:2013wma}
\begin{equation}
\alpha_{s}(\mu_{ij})=\frac{\alpha_{0}}{\ln\left(\frac{\mu_{ij}^{2}+\mu_{0}^{2}}{\Lambda_{0}^{2}} \right)} \,,
\end{equation}
in which $\alpha_{0}$, $\mu_{0}$ and $\Lambda_{0}$ are parameters of the model.

The central terms of Goldstone-boson exchange interaction in CSM can be written as
\begin{align}
&
V_{\pi}\left( \vec{r}_{ij} e^{i\theta} \right) = \frac{g_{ch}^{2}}{4\pi}
\frac{m_{\pi}^2}{12m_{i}m_{j}} \frac{\Lambda_{\pi}^{2}}{\Lambda_{\pi}^{2}-m_{\pi}
^{2}}m_{\pi} \Bigg[ Y(m_{\pi}r_{ij} e^{i\theta}) \nonumber \\
&
\hspace*{1.20cm} - \frac{\Lambda_{\pi}^{3}}{m_{\pi}^{3}}
Y(\Lambda_{\pi}r_{ij} e^{i\theta}) \bigg] (\vec{\sigma}_{i}\cdot\vec{\sigma}_{j})\sum_{a=1}^{3}(\lambda_{i}^{a}
\cdot\lambda_{j}^{a}) \,, \\
& 
V_{\sigma}\left( \vec{r}_{ij} e^{i\theta} \right) = - \frac{g_{ch}^{2}}{4\pi}
\frac{\Lambda_{\sigma}^{2}}{\Lambda_{\sigma}^{2}-m_{\sigma}^{2}}m_{\sigma} \Bigg[
Y(m_{\sigma}r_{ij} e^{i\theta}) \nonumber \\
&
\hspace*{1.20cm} - \frac{\Lambda_{\sigma}}{m_{\sigma}}Y(\Lambda_{\sigma}r_{ij} e^{i\theta})
\Bigg] \,, \\
& 
V_{K}\left( \vec{r}_{ij} e^{i\theta} \right)= \frac{g_{ch}^{2}}{4\pi}
\frac{m_{K}^2}{12m_{i}m_{j}} \frac{\Lambda_{K}^{2}}{\Lambda_{K}^{2}-m_{K}^{2}}m_{
K} \Bigg[ Y(m_{K}r_{ij} e^{i\theta}) \nonumber \\
&
\hspace*{1.20cm} -\frac{\Lambda_{K}^{3}}{m_{K}^{3}}Y(\Lambda_{K}r_{ij} e^{i\theta})
\Bigg] (\vec{\sigma}_{i}\cdot\vec{\sigma}_{j})\sum_{a=4}^{7}(\lambda_{i}^{a}
\cdot\lambda_{j}^{a}) \,, \\
& 
V_{\eta}\left( \vec{r}_{ij} e^{i\theta} \right) = \frac{g_{ch}^{2}}{4\pi}
\frac{m_{\eta}^2}{12m_{i}m_{j}} \frac{\Lambda_{\eta}^{2}}{\Lambda_{\eta}^{2}-m_{
\eta}^{2}}m_{\eta} \Bigg[ Y(m_{\eta}r_{ij} e^{i\theta}) \nonumber \\
&
\hspace*{1.20cm} -\frac{\Lambda_{\eta}^{3}}{m_{\eta}^{3}
}Y(\Lambda_{\eta}r_{ij} e^{i\theta}) \Bigg] (\vec{\sigma}_{i}\cdot\vec{\sigma}_{j})
\Big[\cos\theta_{p} \left(\lambda_{i}^{8}\cdot\lambda_{j}^{8}
\right) \nonumber \\
&
\hspace*{1.20cm} -\sin\theta_{p} \Big] \,,
\end{align}
where $Y(x)=e^{-x}/x$ is the standard Yukawa function. The physical $\eta$ meson are considered by introducing the angle $\theta_p$ instead of the octet one. The $\lambda^{a}$ are the SU(3) flavor Gell-Mann matrices. Taken from their experimental values, $m_{\pi}$, $m_{K}$ and $m_{\eta}$ are the masses of the SU(3) Goldstone bosons. The value of $m_{\sigma}$ is determined through the PCAC relation $m_{\sigma}^{2}\simeq m_{\pi}^{2}+4m_{u,d}^{2}$~\cite{Scadron:1982eg}. Finally, the chiral coupling constant, $g_{ch}$, is determined from the $\pi NN$ coupling constant through
\begin{equation}
\frac{g_{ch}^{2}}{4\pi}=\frac{9}{25}\frac{g_{\pi NN}^{2}}{4\pi} \frac{m_{u,d}^{2}}{m_{N}^2} \,,
\end{equation}
which assumes that flavor SU(3) is an exact symmetry only broken by the different mass of the strange quark.

One need to mention that the chiral quark-(anti)quark interaction only play a role between two light quarks, and it is invalid for the other heavy-light and heavy-heavy quark pairs due to the isospin symmetry breaking. The model parameters which are listed in Table~\ref{model} have been fixed in advance reproducing hadron~\cite{Valcarce:1995dm, Vijande:2004he, Segovia:2008zza, Segovia:2008zz, Ortega:2016hde, Yang:2017xpp}, hadron-hadron ~\cite{Fernandez:1993hx, Valcarce:1994nr, Ortega:2009hj, Ortega:2016mms, Ortega:2016pgg} and multiquark~\cite{Vijande:2006jf, Yang:2015bmv, Yang:2017rpg, gy:2019prdnn} phenomenology.
%For clarity, the ones involved in this calculation are listed in Table~\ref{model}. 

\begin{table}[!t]
\caption{\label{model} Model parameters.}
\begin{ruledtabular}
\begin{tabular}{cccc}
Quark masses     & $m_u=m_d$ (MeV) &  313 \\
                 & $m_c$ (MeV)     & 1752 \\
                 & $m_b$ (MeV)     & 5100 \\[2ex]
Goldstone bosons & $\Lambda_\pi=\Lambda_\sigma~$ (fm$^{-1}$) &   4.20 \\
                 & $\Lambda_\eta$ (fm$^{-1}$)     &   5.20 \\
                 & $g^2_{ch}/(4\pi)$                         &   0.54 \\
                 & $\theta_P(^\circ)$                        & -15 \\[2ex]
Confinement      & $a_c$ (MeV)         & 430\\
                 & $\mu_c$ (fm$^{-1})$ &   0.70\\
                 & $\Delta$ (MeV)      & 181.10 \\[2ex]
                 & $\alpha_0$              & 2.118 \\
                 & $\Lambda_0~$(fm$^{-1}$) & 0.113 \\
OGE              & $\mu_0~$(MeV)        & 36.976\\
                 & $\hat{r}_0~$(MeV~fm) & 28.170\\
\end{tabular}
\end{ruledtabular}
\end{table}

\begin{figure}[ht]
\epsfxsize=3.1in \epsfbox{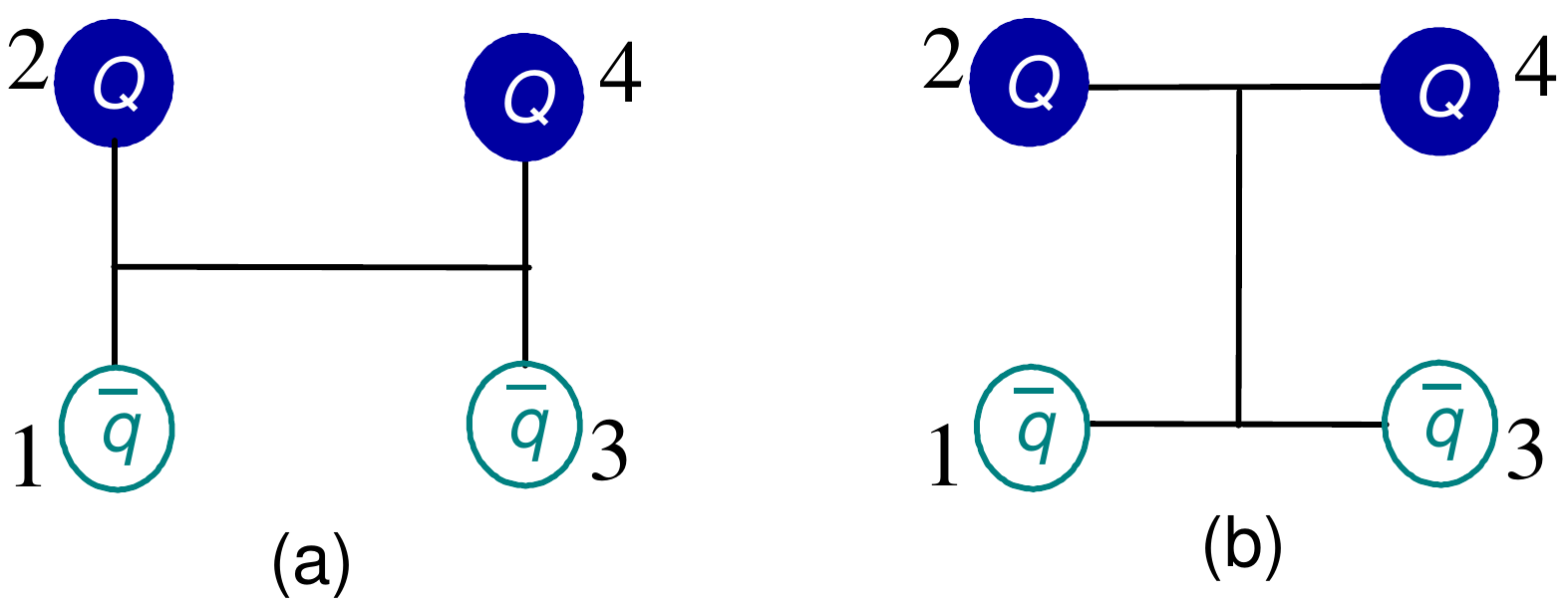}
\caption{Two types of configurations in doubly-heavy tetraquarks. Panel $(a)$ is meson-meson structure and panel $(b)$ is diquark-antidiquark one. $(Q=c,b$ and $q=u,d)$} \label{QQqq}
\end{figure}

Four fundamental degrees of freedom in quark level: color, flavor, spin and space are generally accepted in the QCD theory and the multiquark system wave function is a product of these four terms. In Fig.~\ref{QQqq}, we show two kinds of configurations for doubly-heavy tetraquarks $QQ\bar{q}\bar{q}$ $(q=u,d$ and $Q=c,b)$. In particular, Fig.~\ref{QQqq}(a) is the meson-meson (MM) structure and diquark-antidiquark (DA) one is of Fig.~\ref{QQqq}(b), both of them and their coupling are considered in our investigation.

Concerning the color degree-of-freedom, more richer structures in multiquark system will be discussed than conventional hadrons $(q\bar{q}$ mesons and $qqq$ baryons). The colorless wave function of 4-quark systems in dimeson configuration can be obtained by either a color-singlet or a hidden-color channel or both. However, this is not an unique path for the authors of  Refs.~\cite{Harvey:1980rva, Vijande:2009kj} assert that it is enough to consider the color singlet channel when all possible excited states of a system are included. 
After a comparison, a more economical way of computing through considering all the possible color structures and their coupling is employed. Firstly, in the color $SU(3)$ group, 
the wave functions of color-singlet (two color-singlet clusters coupling, $1\times 1$) and hidden-color (two color-octet clusters coupling, $8\times 8$) channel in dimeson configuration of Fig.~\ref{QQqq}(a) is signed as $\chi^c_1$ and $\chi^c_2$ respectively,
\begin{align}
\label{Color1}
\chi^c_1 &= \frac{1}{3}(\bar{r}r+\bar{g}g+\bar{b}b)\times (\bar{r}r+\bar{g}g+\bar{b}b) \,,
\end{align}

\begin{align}
\label{Color2}
\chi^c_2 &= \frac{\sqrt{2}}{12}(3\bar{b}r\bar{r}b+3\bar{g}r\bar{r}g+3\bar{b}g\bar{g}b+3\bar{g}b\bar{b}g+3\bar{r}g\bar{g}r
\nonumber\\
&+3\bar{r}b\bar{b}r+2\bar{r}r\bar{r}r+2\bar{g}g\bar{g}g+2\bar{b}b\bar{b}b-\bar{r}r\bar{g}g
\nonumber\\
&-\bar{g}g\bar{r}r-\bar{b}b\bar{g}g-\bar{b}b\bar{r}r-\bar{g}g\bar{b}b-\bar{r}r\bar{b}b) \,.
\end{align}
In additional, also according to an increased sequence of numbers labeled in Fig.~\ref{QQqq}, the color wave functions of diquark-antidiquark structure shown in Fig.~\ref{QQqq}(b) are $\chi^c_3$ (color triplet-antitriplet clusters coupling, $3\times \bar{3}$) and $\chi^c_4$ (color sextet-antisextet clusters coupling, $6\times \bar{6}$), respectively:
\begin{align}
\label{Color3}
\chi^c_3 &= \frac{\sqrt{3}}{6}(\bar{r}r\bar{g}g-\bar{g}r\bar{r}g+\bar{g}g\bar{r}r-\bar{r}g\bar{g}r+\bar{r}r\bar{b}b
\nonumber\\
&-\bar{b}r\bar{r}b+\bar{b}b\bar{r}r-\bar{r}b\bar{b}r+\bar{g}g\bar{b}b-\bar{b}g\bar{g}b
\nonumber\\
&+\bar{b}b\bar{g}g-\bar{g}b\bar{b}g) \,,
\end{align}
\begin{align}
\label{Color4}
\chi^c_4 &= \frac{\sqrt{6}}{12}(2\bar{r}r\bar{r}r+2\bar{g}g\bar{g}g+2\bar{b}b\bar{b}b+\bar{r}r\bar{g}g+\bar{g}r\bar{r}g
\nonumber\\
&+\bar{g}g\bar{r}r+\bar{r}g\bar{g}r+\bar{r}r\bar{b}b+\bar{b}r\bar{r}b+\bar{b}b\bar{r}r
\nonumber\\
&+\bar{r}b\bar{b}r+\bar{g}g\bar{b}b+\bar{b}g\bar{g}b+\bar{b}b\bar{g}g+\bar{g}b\bar{b}g) \,.
\end{align}

As for the flavor degree-of freedom, due to the quark contents of the present investigated 4-quark systems are two heavy quarks $(Q=c,d)$ and two light antiquarks $(\bar{q}=\bar{u},\bar{d})$, only isospin $I=0$ and $1$ will be obtained. Moreover, the flavor wave-functions signed as $\chi^{fi}_{I, M_I}$ with the superscript $i=1,~2$ and $3$ are of $cc\bar{q}\bar{q}$, $bb\bar{q}\bar{q}$ and $cb\bar{q}\bar{q}$ systems, respectively. The specific wave functions read as below,
\begin{align}
&
\chi_{0,0}^{f1} = \sqrt{\frac{1}{2}} (\bar{u}c\bar{d}c-\bar{d}c\bar{u}c) \,, \\
&
\chi_{1,-1}^{f1} = \bar{u}c\bar{u}c \,, \\
&
\chi_{0,0}^{f2} = \sqrt{\frac{1}{2}} (\bar{u}b\bar{d}b-\bar{d}b\bar{u}b) \,, \\
&
\chi_{1,-1}^{f2} = \bar{u}b\bar{u}b \,, \\
&
\chi_{0,0}^{f3} = \sqrt{\frac{1}{2}} (\bar{u}c\bar{d}b-\bar{d}c\bar{u}b) \,,  \\
&
\chi_{1,-1}^{f3} = \bar{u}c\bar{u}b \,,
\end{align}
where the third component of the isospin $M_I$ is set to be equal to the absolute value of total one $I$ without loss of generality for there is no interplay in the Hamiltonian that can distinguish such component.

We consider herein 4-quark bound states with total spin $S$ ranging from $0$ to $2$. Since there is not any spin-orbital coupling dependent potential included in our Hamiltonian, the third component $(M_S)$ of tetraquark spin can be assumed to be equal to the total one without loss of generality too. Our total spin wave functions $\chi^{\sigma}_{S, M_S}$ are given by:
\begin{align}
\label{Spin}
\chi_{0,0}^{\sigma 1}(4) &= \chi^\sigma_{00}\chi^\sigma_{00} \\
\chi_{0,0}^{\sigma 2}(4) &= \frac{1}{\sqrt{3}}(\chi^\sigma_{11}\chi^\sigma_{1,-1}-\chi^\sigma_{10}\chi^\sigma_{10}+\chi^\sigma_{1,-1}\chi^\sigma_{11}) \\
\label{SWF1}
\chi_{1,1}^{\sigma 1}(4) &= \chi^\sigma_{00}\chi^\sigma_{11} \\ 
\label{SWF2}
\chi_{1,1}^{\sigma 2}(4) &= \chi^\sigma_{11}\chi^\sigma_{00} \\
\chi_{1,1}^{\sigma 3}(4) &= \frac{1}{\sqrt{2}} (\chi^\sigma_{11} \chi^\sigma_{10}-\chi^\sigma_{10} \chi^\sigma_{11}) \\
\chi_{2,2}^{\sigma 1}(4) &= \chi^\sigma_{11}\chi^\sigma_{11} 
\end{align}
these expressions are obtained by considering the coupling of two sub-clusters spin wave functions with SU(2) algebra, and the necessary bases are read as
\begin{align}
\label{Spin}
\chi^\sigma_{11} &= \alpha \alpha \,,  \chi^\sigma_{1,-1} = \beta \beta \\
\chi^\sigma_{10} &= \frac{1}{\sqrt{2}}(\alpha \beta+\beta \alpha) \\
\chi^\sigma_{00} &= \frac{1}{\sqrt{2}}(\alpha \beta-\beta \alpha) 
\end{align}

Here, one important thing need to be mentioned that the spin wave functions of Eq.~(\ref{SWF1}) and~(\ref{SWF2}) are equivalent for two $D$- or $B$-mesons configuration of tetraquark state. Namely, the calculated masses of $DD^*$ and $D^*D$ are exactly the same (also for $BB^*$) and obviously, this is a trivial fact in hadron level.

Among the different methods to solve the Schr\"odinger-like 4-body bound state equation, we use the Rayleigh-Ritz variational principle which is one of the most extended tools to solve eigenvalue problems due to its simplicity and flexibility. Meanwhile, the choice of basis to expand the intrinsic wave function of state is of great importance. In the relative motion coordinates of 4-quark systems, the spatial wave function is written as follows:
\begin{equation}
\label{eq:WFexp}
\psi_{LM_L}(\theta)= \left[ \left[ \phi_{n_1l_1}(\vec{\rho}e^{i\theta}\,) \phi_{n_2l_2}(\vec{\lambda}e^{i\theta}\,)\right]_{l} \phi_{n_3l_3}(\vec{R}e^{i\theta}\,) \right]_{L M_L} \,,
\end{equation}
where the internal Jacobi coordinates for Fig.~\ref{QQqq}(a) of meson-meson configuration are defined as
\begin{align}
\vec{\rho} &= \vec{x}_1-\vec{x}_2 \,, \\
\vec{\lambda} &= \vec{x}_3 - \vec{x}_4 \,, \\
\vec{R} &= \frac{m_1 \vec{x}_1 + m_2 \vec{x}_2}{m_1+m_2}- \frac{m_3 \vec{x}_3 + m_4 \vec{x}_4}{m_3+m_4} \,,
\end{align}
and the diquark-antdiquark structure of Fig.~\ref{QQqq}(b) are,
\begin{align}
\vec{\rho} &= \vec{x}_1-\vec{x}_3 \,, \\
\vec{\lambda} &= \vec{x}_2 - \vec{x}_4 \,, \\
\vec{R} &= \frac{m_1 \vec{x}_1 + m_3 \vec{x}_3}{m_1+m_3}- \frac{m_2 \vec{x}_2 + m_4 \vec{x}_4}{m_2+m_4} \,.
\end{align}
Obviously, with these sets of coordinates the center-of-mass kinetic term $T_{CM}$ can be completely eliminated for a nonrelativistic system. Besides, the Jacobi coordinates of Eq.~(\ref{eq:WFexp}) are also transformed with a common scaling angle $\theta$.

A high efficiency and exact method in solving bound state of few-body system, Gaussian expansion method (GEM)~\cite{Hiyama:2003cu} is employed in this work, all of the relative motions of 4-quark systems are expanded with various Gaussian basis which are taken as the geometric progression sizes\footnote{The details on Gaussian parameters can be found in Ref.~\cite{Yang:2015bmv}}, and the form of orbital wave functions, $\phi$'s in Eq.~\eqref{eq:WFexp} is 
\begin{align}
&
\phi_{nlm}(\vec{r}e^{i\theta}\,) = N_{nl} (re^{i\theta})^{l} e^{-\nu_{n} (re^{i\theta})^2} Y_{lm}(\hat{r}) \,.
%&
%= N_{nl} \lim_{\varepsilon\to 0} \frac{1}{(\nu_{n}\varepsilon)^l} \sum_{k=1}^{k_{\rm
%max}} C_{lm,k} e^{-\nu_{n}(\vec{r}-\varepsilon \vec{D}_{lm,k})^{2}} \,.
\end{align}
Moreover, our present study is only in $S$-wave state of doubly-heavy tetraquarks, no laborious Racah algebra during matrix elements calculation for the value of spherical harmonic function is a constant when $l=0$, $i. e.$ $Y_{00}=\sqrt{1/4\pi}$.

Finally, in order to fulfill the Pauli principle, the complete wave-function is written as
\begin{equation}
\label{TPs}
\Psi_{JM_J,I,i,j,k}(\theta)={\cal A} \left[ \left[ \psi_{L}(\theta) \chi^{\sigma_i}_{S}(4) \right]_{JM_J} \chi^{f_j}_I \chi^{c}_k \right] \,,
\end{equation}
where $\cal{A}$ is the antisymmetry operator of doubly-heavy tetraquarks by considering the nature of identical particle interchange $(\bar{q}\bar{q},~cc$ and $bb)$. This is necessary for the complete wave function of the 4-quark system is constructed from two sub-clusters, $i. e.$ meson-meson and diquark-antidiquark structures. 
In particular, when the two heavy quarks are of the same flavor $(QQ=cc$ or $bb)$, the definitions of these two configurations in Fig.~\ref{QQqq} with the quark arrangements of $\bar{q}Q\bar{q}Q$ are both
\begin{equation}
{\cal{A}} = 1-(13)-(24)+(13)(24) \,.
\end{equation}
However, due to the asymmetry between $c$- and $b$-quark, it is only two terms for $\bar{q}c\bar{q}b$ system and read as
\begin{equation}
{\cal{A}} = 1-(13) \,.
\end{equation}

%%%%%%%%%%%%%%%%%%%%%%%%%%%%%%%%%%%%%%%%%%%%%%%%%%%%%%%%%%%%%%%%%%%%%%%%%%%%%%%%%%%%%%%%%%

\begin{table*}[!t]
\caption{\label{GDD} All possible channels for $cc\bar{q}\bar{q}$ $(q=u~or~d)$ tetraquark systems.}
\begin{ruledtabular}
\begin{tabular}{cccccc}
& & \multicolumn{2}{c}{$I=0$} & \multicolumn{2}{c}{$I=1$} \\[2ex]
$J^P$~~&~Index~ & $\chi_J^{\sigma_i}$;~$\chi_I^{f_j}$;~$\chi_k^c$ & Channel~~ & $\chi_J^{\sigma_i}$;~$\chi_I^{f_j}$;~$\chi_k^c$ & Channel~~ \\
&&$[i; ~j; ~k]$& &$[i; ~j; ~k]$&  \\[2ex]
$0^+$ & 1  & $[1; ~1; ~1]$   & $(D^+ D^0)^1$ & $[1; ~1; ~1]$   & $(D^0 D^0)^1$ \\
&  2 & $[2; ~1; ~1]$ & $(D^{*+} D^{*0})^1$ & $[2; ~1; ~1]$  & $(D^{*0} D^{*0})^1$ \\
&  3 & $[1; ~1; ~2]$   & $(D^+ D^0)^8$  & $[1; ~1; ~2]$   & $(D^0 D^0)^8$ \\
&  4 & $[2; ~1; ~2]$ & $(D^{*+} D^{*0})^8$ & $[2; ~1; ~2]$   & $(D^{*0} D^{*0})^8$   \\
&  5  &                     &                              & $[3; ~1; ~4]$   & $(cc)(\bar{u}\bar{u})$ \\
&  6 &                      &                              & $[4; ~1; ~3]$   & $(cc)^*(\bar{u}\bar{u})^*$ \\[2ex]
$1^+$ & 1  & $[1; ~1; ~1]$   & $(D^+ D^{*0})^1$ & $[1; ~1; ~1]$   & $(D^0 D^{*0})^1$\\
%& 2  & $[2; ~1; ~1]$  & $(D^{*+} D^0)^1$ & $[2; ~1; ~1]$  & $(D^{*0} D^0)^1$ \\
& 2  & $[3; ~1; ~1]$     & $(D^{*+} D^{*0})^1$  & $[3; ~1; ~1]$   & $(D^{*0} D^{*0})^1$ \\
& 3  & $[1; ~1; ~2]$  & $(D^+ D^{*0})^8$ & $[1; ~1; ~2]$ & $(D^0 D^{*0})^8$ \\
%& 4  & $[2; ~1; ~2]$     & $(D^{*+} D^0)^8$  & $[2; ~1; ~2]$   & $(D^{*0} D^0)^8$ \\
& 4  & $[3; ~1; ~2]$  & $(D^{*+} D^{*0})^8$  & $[3; ~1; ~2]$   & $(D^{*0} D^{*0})^8$  \\
& 5  & $[4; ~1; ~3]$     & $(cc)^*(\bar{u}\bar{d})$  & $[6; ~1; ~3]$   & $(cc)^*(\bar{u}\bar{u})^*$ \\
& 6  & $[5; ~1; ~4]$  & $(cc)(\bar{u}\bar{d})^*$  &                          &   \\[2ex]
$2^+$ & 1  & $[1; ~1; ~1]$   & $(D^{*+} D^{*0})^1$ & $[1; ~1; ~1]$   & $(D^{*0} D^{*0})^1$ \\
& 2  & $[1; ~1; ~2]$  & $(D^{*+} D^{*0})^8$ & $[1; ~1; ~2]$  & $(D^{*0} D^{*0})^8$ \\
& 3  &   &   & $[1; ~1; ~3]$   & $(cc)^*(\bar{u}\bar{u})^*$ \\
\end{tabular}
\end{ruledtabular}
\end{table*}

\begin{table*}[!t]
\caption{\label{GBB} All possible channels for $bb\bar{q}\bar{q}$ $(q=u~or~d)$ tetraquark systems.}
\begin{ruledtabular}
\begin{tabular}{cccccc}
& & \multicolumn{2}{c}{$I=0$} & \multicolumn{2}{c}{$I=1$} \\[2ex]
$J^P$~~&~Index~ & $\chi_J^{\sigma_i}$;~$\chi_I^{f_j}$;~$\chi_k^c$ & Channel~~ & $\chi_J^{\sigma_i}$;~$\chi_I^{f_j}$;~$\chi_k^c$ & Channel~~ \\
&&$[i; ~j; ~k]$& &$[i; ~j; ~k]$&  \\[2ex]
$0^+$ & 1  & $[1; ~2; ~1]$   & $(B^- \bar{B}^0)^1$ & $[1; ~2; ~1]$   & $(B^- B^-)^1$ \\
&  2 & $[2; ~2; ~1]$ & $(B^{*-} \bar{B}^{*0})^1$ & $[2; ~2; ~1]$  & $(B^{*-} B^{*-})^1$ \\
&  3 & $[1; ~2; ~2]$   & $(B^- \bar{B}^0)^8$  & $[1; ~2; ~2]$   & $(B^- B^-)^8$ \\
&  4 & $[2; ~2; ~2]$ & $(B^{*-} \bar{B}^{*0})^8$ & $[2; ~2; ~2]$   & $(B^{*-} B^{*-})^8$   \\
&  5  &                     &                              & $[3; ~2; ~4]$   & $(bb)(\bar{u}\bar{u})$ \\
&  6 &                      &                              & $[4; ~2; ~3]$   & $(bb)^*(\bar{u}\bar{u})^*$ \\[2ex]
$1^+$ & 1  & $[1; ~2; ~1]$   & $(B^- \bar{B}^{*0})^1$ & $[1; ~2; ~1]$   & $(B^- B^{*-})^1$\\
%& 2  & $[2; ~2; ~1]$  & $(B^{*-} \bar{B}^0)^1$ & $[2; ~2; ~1]$  & $(B^{*-} B^-)^1$ \\
& 2  & $[3; ~2; ~1]$     & $(B^{*-} \bar{B}^{*0})^1$  & $[3; ~2; ~1]$   & $(B^{*-} B^{*-})^1$ \\
& 3  & $[1; ~2; ~2]$  & $(B^- \bar{B}^{*0})^8$ & $[1; ~2; ~2]$ & $(B^- B^{*-})^8$ \\
%& 5  & $[2; ~2; ~2]$     & $(B^{*-} \bar{B}^0)^8$  & $[2; ~2; ~2]$   & $(B^{*-} B^-)^8$ \\
& 4  & $[3; ~2; ~2]$  & $(B^{*-} \bar{B}^{*0})^8$  & $[3; ~2; ~2]$   & $(B^{*-} B^{*-})^8$  \\
& 5  & $[4; ~2; ~3]$     & $(bb)^*(\bar{u}\bar{d})$  & $[6; ~2; ~3]$   & $(bb)^*(\bar{u}\bar{u})^*$ \\
& 6  & $[5; ~2; ~4]$  & $(bb)(\bar{u}\bar{d})^*$  &                          &   \\[2ex]
$2^+$ & 1  & $[1; ~2; ~1]$   & $(B^{*-} \bar{B}^{*0})^1$ & $[1; ~2; ~1]$   & $(B^{*-} B^{*-})^1$ \\
& 2  & $[1; ~2; ~2]$  & $(B^{*-} \bar{B}^{*0})^8$ & $[1; ~2; ~2]$  & $(B^{*-} B^{*-})^8$ \\
& 3  &   &   & $[1; ~2; ~3]$   & $(bb)^*(\bar{u}\bar{u})^*$ \\
\end{tabular}
\end{ruledtabular}
\end{table*}

\begin{table*}[!t]
\caption{\label{GDB} All possible channels for $cb\bar{q}\bar{q}$ $(q=u~or~d)$ tetraquark systems. For a brief purpose, only the $D^{(*)0} B^{(*)0}$ structures are listed and the corresponding $D^{(*)+} \bar{B}^{(*)-}$ ones are absent in $I=0$. However, all these configurations are still employed in constructing the wavefunctions of 4-quark systems.}
\begin{ruledtabular}
\begin{tabular}{cccccc}
& & \multicolumn{2}{c}{$I=0$} & \multicolumn{2}{c}{$I=1$} \\[2ex]
$J^P$~~&~Index~ & $\chi_J^{\sigma_i}$;~$\chi_I^{f_j}$;~$\chi_k^c$ & Channel~~ & $\chi_J^{\sigma_i}$;~$\chi_I^{f_j}$;~$\chi_k^c$ & Channel~~ \\
&&$[i; ~j; ~k]$& &$[i; ~j; ~k]$&  \\[2ex]
$0^+$ & 1  & $[1; ~3; ~1]$   & $(D^0 \bar{B}^0)^1$ & $[1; ~3; ~1]$   & $(D^0 B^-)^1$ \\
&  2 & $[2; ~3; ~1]$ & $(D^{*0} \bar{B}^{*0})^1$ & $[2; ~3; ~1]$  & $(D^{*0} B^{*-})^1$ \\
&  3 & $[1; ~3; ~2]$   & $(D^0 \bar{B}^0)^8$  & $[1; ~3; ~2]$   & $(D^0 B^-)^8$ \\
&  4 & $[2; ~3; ~2]$ & $(D^{*0} \bar{B}^{*0})^8$ & $[2; ~3; ~2]$   & $(D^{*0} B^{*-})^8$   \\
&  5  & $[3; ~3; ~3]$ & $(cb)(\bar{u}\bar{d})$  & $[3; ~3; ~4]$   & $(cb)(\bar{u}\bar{u})$ \\
&  6 & $[4; ~3; ~4]$  & $(cb)^*(\bar{u}\bar{d})^*$ & $[4; ~3; ~3]$   & $(cb)^*(\bar{u}\bar{u})^*$ \\[2ex]
$1^+$ & 1  & $[1; ~3; ~1]$   & $(D^0 \bar{B}^{*0})^1$ & $[1; ~3; ~1]$   & $(D^0 B^{*-})^1$\\
& 2  & $[2; ~3; ~1]$  & $(D^{*0} \bar{B}^0)^1$ & $[2; ~3; ~1]$  & $(D^{*0} B^-)^1$ \\
& 3  & $[3; ~3; ~1]$     & $(D^{*0} \bar{B}^{*0})^1$  & $[3; ~3; ~1]$   & $(D^{*0} B^{*-})^1$ \\
& 4  & $[1; ~3; ~2]$  & $(D^0 \bar{B}^{*0})^8$ & $[1; ~3; ~2]$ & $(D^0 B^{*-})^8$ \\
& 5  & $[2; ~3; ~2]$     & $(D^{*0} \bar{B}^0)^8$  & $[2; ~3; ~2]$   & $(D^{*0} B^-)^8$ \\
& 6  & $[3; ~3; ~2]$  & $(D^{*0} \bar{B}^{*0})^8$  & $[3; ~3; ~2]$   & $(D^{*0} B^{*-})^8$  \\
& 7  & $[4; ~3; ~3]$     & $(cb)^*(\bar{u}\bar{d})$  & $[4; ~3; ~4]$   & $(cb)^*(\bar{u}\bar{u})$ \\
& 8  & $[5; ~3; ~4]$  & $(cb)(\bar{u}\bar{d})^*$  &  $[5; ~3; ~3]$    & $(cb)(\bar{u}\bar{u})^*$   \\
& 9  & $[6; ~3; ~4]$  & $(cb)^*(\bar{u}\bar{d})^*$  &  $[6; ~3; ~3]$    & $(cb)^*(\bar{u}\bar{u})^*$   \\[2ex]
$2^+$ & 1  & $[1; ~3; ~1]$   & $(D^{*0} \bar{B}^{*0})^1$ & $[1; ~3; ~1]$   & $(D^{*0} B^{*-})^1$ \\
& 2  & $[1; ~3; ~2]$  & $(D^{*0} \bar{B}^{*0})^8$ & $[1; ~3; ~2]$  & $(D^{*0} B^{*-})^8$ \\
& 3  & $[1; ~3; ~4]$   & $(cb)^*(\bar{u}\bar{d})^*$ & $[1; ~3; ~3]$   & $(cb)^*(\bar{u}\bar{u})^*$ \\
\end{tabular}
\end{ruledtabular}
\end{table*}

\section{Results}
\label{sec:results}

In the present work, we systematically investigate the low-lying $S$-wave states of $QQ\bar{q}\bar{q}$ $(q=u,c$ and $Q=c,b)$ tetraquarks which both meson-meson and diquark-antidiquark configurations are considered. The parity for different doubly-heavy tetraquarks is positive under our assumption that the angular momenta $l_1$, $l_2$, $l_3$, which appear in Eq.~\eqref{eq:WFexp}, are all $0$. In this way, the total angular momentum, $J$, coincides with the total spin, $S$, and can take values of $0$, $1$ and $2$. All possible dimeson and diquark-antidiquark channels for $cc\bar{q}\bar{q}$, $bb\bar{q}\bar{q}$ and $cb\bar{q}\bar{q}$ systems are listed in Table~\ref{GDD}, \ref{GBB} and \ref{GDB} respectively, and they have been grouped according to total spin-pairty $J^P$ and isospin $I$. For a clarity purpose, the third and fifth columns of these tables show the necessary basis combination in spin $(\chi^{\sigma_i}_J)$, flavor $(\chi^{f_j}_I)$, and color $(\chi^c_k)$ degrees-of-freedom. The physical channels with color-singlet (labeled with the superindex $1$), hidden-color (labeled with the superindex $8$) and diquark-antidiquark (labeled with $(QQ)(\bar{q}\bar{q})$) configurations are listed in the fourth and sixth columns. 

Tables range from~\ref{GresultCC1} to~\ref{SumRest} summarized our calculated results (mass, size and component) of possible lowest-lying doubly-heavy tetraquarks. In particular, Tables~\ref{GresultCompDD},~\ref{GresultCompBB} and~\ref{GresultCompDB1} list each components of possible bound states of doubly-charm, doubly-bottom and charm-bottom tetraquarks in the complete coupled-channels calculation which all possible channels for a given quantum number $IJ^P$ are considered. Their inner structures, the distance among any quark pair is shown in Tables~\ref{tab:disDD},~\ref{tab:disBB} and~\ref{tab:disDB1}, this is in order to get some insight about either molecular or compact tetraquark we are dealing with. The rest tables below are of the calculated masses of these bound or resonance states of doubly-heavy tetraquarks, namely Tables~\ref{GresultCC1} and~\ref{GresultBB1} present the results of doubly-charm and doubly-bottom tetraquarks which quantum numbers are both of $I(J^P)=0(1^+)$, and results on charm-bottom tetraquarks with $I(J^P)=0(0^+)$, $0(1^+)$ and $0(2^+)$ are in Tables~\ref{GresultCB1},~\ref{GresultCB2} and~\ref{GresultCB3} respectively. Table~\ref{SumRest} summarizes the obtained bound and resonance states of doubly-heavy tetraquarks in the complete coupled-channels calculation. Moreover, Fig.~\ref{PP1} to Fig.~\ref{PP5} present the distribution of complex energies of these doubly-heavy tetraquarks in coupled-channels calculation by complex scaling method. The transverse direction is of the real part of complex energy $E$, it stands for the mass of tetraquarks, and the longitudinal one is the imaginary part of $E$ which is related to the width, $\Gamma=-2Im(E)$.
However, the other quantum states of each doubly-heavy tetraquarks sectors do not appear here also have been considered in the calculation but neither bound nor resonance states are found. 

 In Tables~\ref{GresultCC1},~\ref{GresultBB1},~\ref{GresultCB1},~\ref{GresultCB2} and~\ref{GresultCB3}, the first column lists the physical channel of meson-meson and diquark-antidiquark (if it fulfills Pauli principle), and the experimental value of the noninteracting meson-meson threshold is also indicated in parenthesis; the second column refers to color-singlet (S), hidden-color (H) and coupled-channels (S+H) calculations for meson-meson configuration; the following two columns show the theoretical mass $(M)$ and binding energy $(E_B)$ of tetraquark state; moreover, as to avoid theoretical uncertainties coming from the quark model prediction of the meson spectra, the last column presents the re-scaled theoretical mass $(M^\prime)$ of tetraquark state by attending to the corresponding experimental meson-meson threshold. 

Now let us proceed to describe in detail our theoretical findings for each sector of doubly-heavy tetraquarks:

% Here need to point out that due to the first two identical quarks in three quarks sub-cluster are symmetry ($l=0$) in spatial part, based on the Pauli exclusion principle, it should be  anti-symmetry after coupling the other three spaces for these two quarks.

\begin{table}[!t]
\caption{\label{GresultCC1} Lowest-lying states of doubly-charm tetraquarks with quantum numbers $I(J^P)=0(1^+)$, unit in MeV.}
\begin{ruledtabular}
\begin{tabular}{lcccc}
Channel   & Color & $M$ & $E_B$ & $M'$ \\[2ex]
$D^+ D^{*0}$ & S   & $3915$ & $0$  & $3877$ \\
$(3877)$           & H   & $4421$ & $+506$ & $4383$ \\
                    & S+H & $3914$ & $-1$  & $3876$ \\
                    & \multicolumn{4}{c}{Percentage (S;H): 97.3\%; 2.7\%} \\[2ex]
%
%$D^*D$ & S   & $11097$ & $-15$ & $11094$ \\
%$(3877)$           & H   & $11175$ & $+63$ & $11172$ \\
%                    & S+H & $11045$ & $-67$ & $11042$ \\
%                    & \multicolumn{4}{c}{Percentage (S;H): 55.5\%; 44.5\%} \\[2ex]
%
$D^{*+} D^{*0}$ & S   & $4034$ & $0$  & $4018$ \\
$(4018)$             & H   & $4390$ & $+356$ & $4374$ \\
                      & S+H & $4033$ & $-1$ & $4017$ \\
                      & \multicolumn{4}{c}{Percentage (S;H): 95.5\%; 4.5\%} \\[2ex] 
$(cc)^*(\bar{u}\bar{d})$ &    & $3778$ &  & \\[2ex] 
$(cc)(\bar{u}\bar{d})^*$ &    & $4220$ &  & \\[2ex]            
Mixed  & & $3726$ & & \\
\end{tabular}
\end{ruledtabular}
\end{table}

\begin{table}[!t]
\caption{{Component of each channel in coupled-channels calculation with $IJ^P=01^+$, the numbers $1$ and $8$ of superscript are for singlet-color and hidden-color channel respectively.}  \label{GresultCompDD}}
\begin{ruledtabular}
\begin{tabular}{lccc}
  ~~$(D^+ D^{*0})^1$~~  & ~~$(D^{*+} D^{*0})^1$~~   & ~~$(D^+ D^{*0})^8$~~ &
   ~~$(D^{*+} D^{*0})^8$~~ \\
 ~~25.8\%~~  & ~~15.4\%~~  & 10.7\%  & 11.2\% ~~\\[2ex]
  ~~$(cc)^*(\bar{u}\bar{d})$~~ & ~~$(cc)(\bar{u}\bar{d})^*$~~  \\ 
 ~~36.7\%  & 0.2\% \\
\end{tabular}
\end{ruledtabular}
\end{table}

\begin{table}[!t]
\caption{\label{tab:disDD} The distance, in fm, between any two quarks of the found tetraquark bound-states in coupled-channels calculation $(q=u, d)$.}
\begin{ruledtabular}
\begin{tabular}{ccc}
  $r_{\bar{u}\bar{d}}$ & $r_{\bar{q}c}$ & $r_{cc}$  \\[2ex]
  0.658 & 0.666 & 0.522 \\
\end{tabular}
\end{ruledtabular}
\end{table}

\subsection{doubly-charm tetraquarks}
In this sector, bound state and resonance are only found in the $I(J^P)=0(1^+)$ state.
%{\bf The $\bm{I(J^P)=0(1^+)}$ channel:}
Two possible meson-meson channels, $D^+ D^{*0}$ and $D^{*+} D^{*0}$, along with two diquark-antidiquark channels, $(cc)^*(\bar{u}\bar{d})$ and $(cc)(\bar{u}\bar{d})^*$ are studied in Table~\ref{GresultCC1}. It is obviously to notice that there is no bound state in neither color-singlet (S) nor hidden-color channels (H) of the meson-meson configuration. However, this result is reversed by their coupled-channels calculation (S+H) and there are $-1\,\text{MeV}$ weakly binding energies both for $D^+ D^{*0}$ and $D^{*+} D^{*0}$ channels. After corrections, the re-scaled masses of these two channels are $3876\,\text{MeV}$ and $4017\,\text{MeV}$, respectively. Meanwhile, the nature of molecular-type $D^{(*)+} D^{*0}$ structures are shown up since the color-singlet channels contributions are more than 95\%. 

In contrast to the weakly bound states around the $D^{(*)+} D^{*0}$ thresholds, there are almost $-140\,\text{MeV}$ binding energy for $(cc)^*(\bar{u}\bar{d})$ channel when compared with the theoretical threshold of $D^+ D^{*0}$. However, the other diquark-antidiquark channel $(cc)(\bar{u}\bar{d})^*$ is above the $D^+ D^{*0}$ and $D^{*+} D^{*0}$ theoretical thresholds with $E_B=+305\,\text{MeV}$ and $+186\,\text{MeV}$, respectively. This deeply bound diquark-antidiquark state $(cc)^*(\bar{u}\bar{d})$ motivates a further complete coupled-channels calculation which all the color-singlet, hidden-color of meson-meson channels and diquark-antidiquark ones are considered. The obtained mass is $3726\,\text{MeV}$ which is $52\,\text{MeV}$ lower than the single channel result of $(cc)^*(\bar{u}\bar{d})$, besides its nature of compact doubly-charm tetraquark state is clearly presented in Table~\ref{tab:disDD} where the distance between any two quarks  are calculated and the obtained size of this four-quark system is less than $0.67\,\text{fm}$. Table~\ref{GresultCompDD} shows each component in the coupled-channels calculation. In particular, two mainly comparable components, color-singlet channel $D^+ D^{*0}$ (25.8\%) and $(cc)^*(\bar{q}\bar{q})$ one (36.7\%), consist with our result of strong coupling effect and compact tetraquark structure. 

\begin{figure}[ht]
\epsfxsize=3.5in \epsfbox{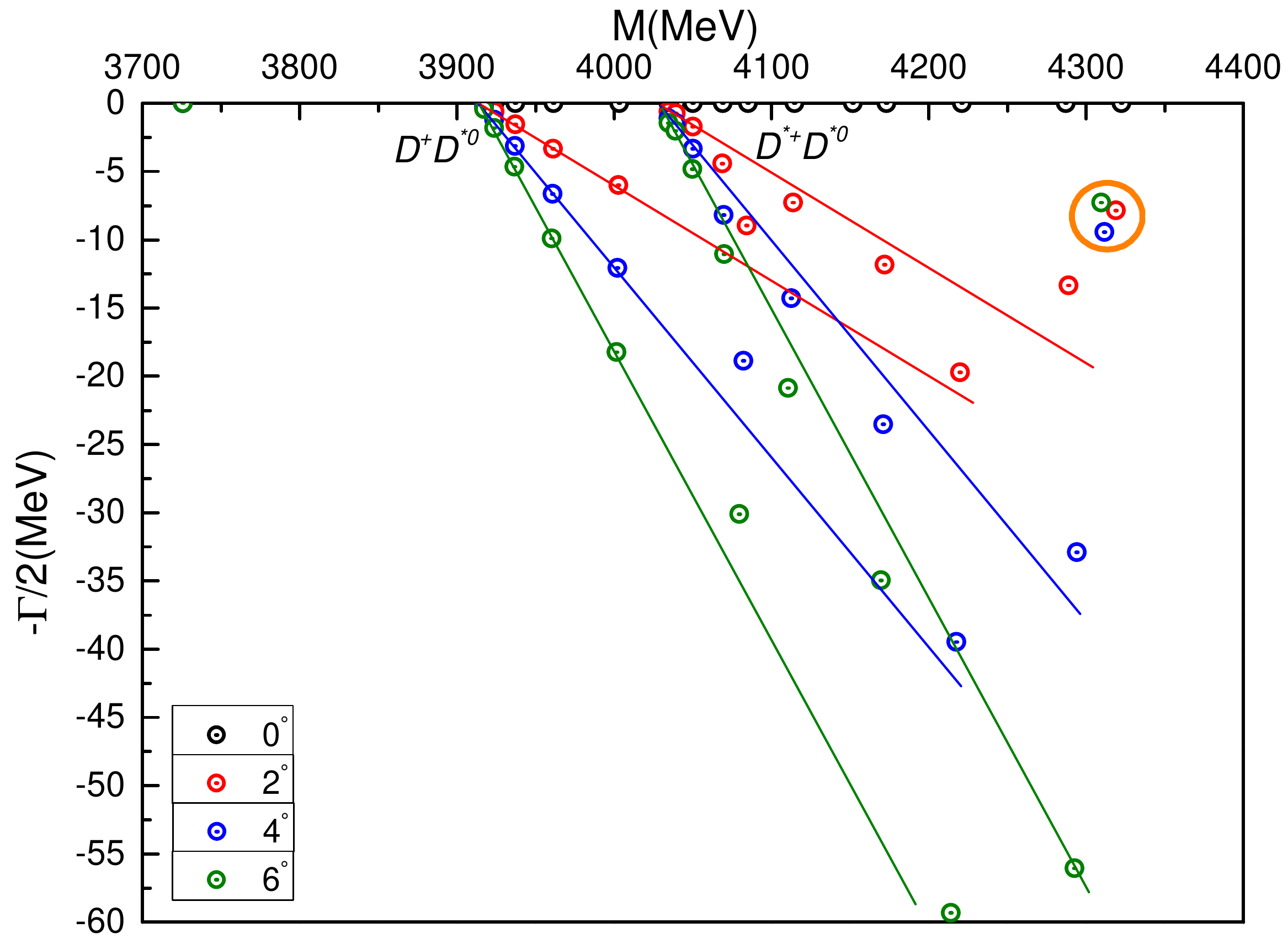}
\caption{Complex energies of doubly-charm tetraquarks with $IJ^P=01^+$ in the coupled channels calculation, $\theta$ varying from $0^\circ$ to $6^\circ$ .} \label{PP1}
\end{figure}

\begin{figure}[ht]
\epsfxsize=3.45in \epsfbox{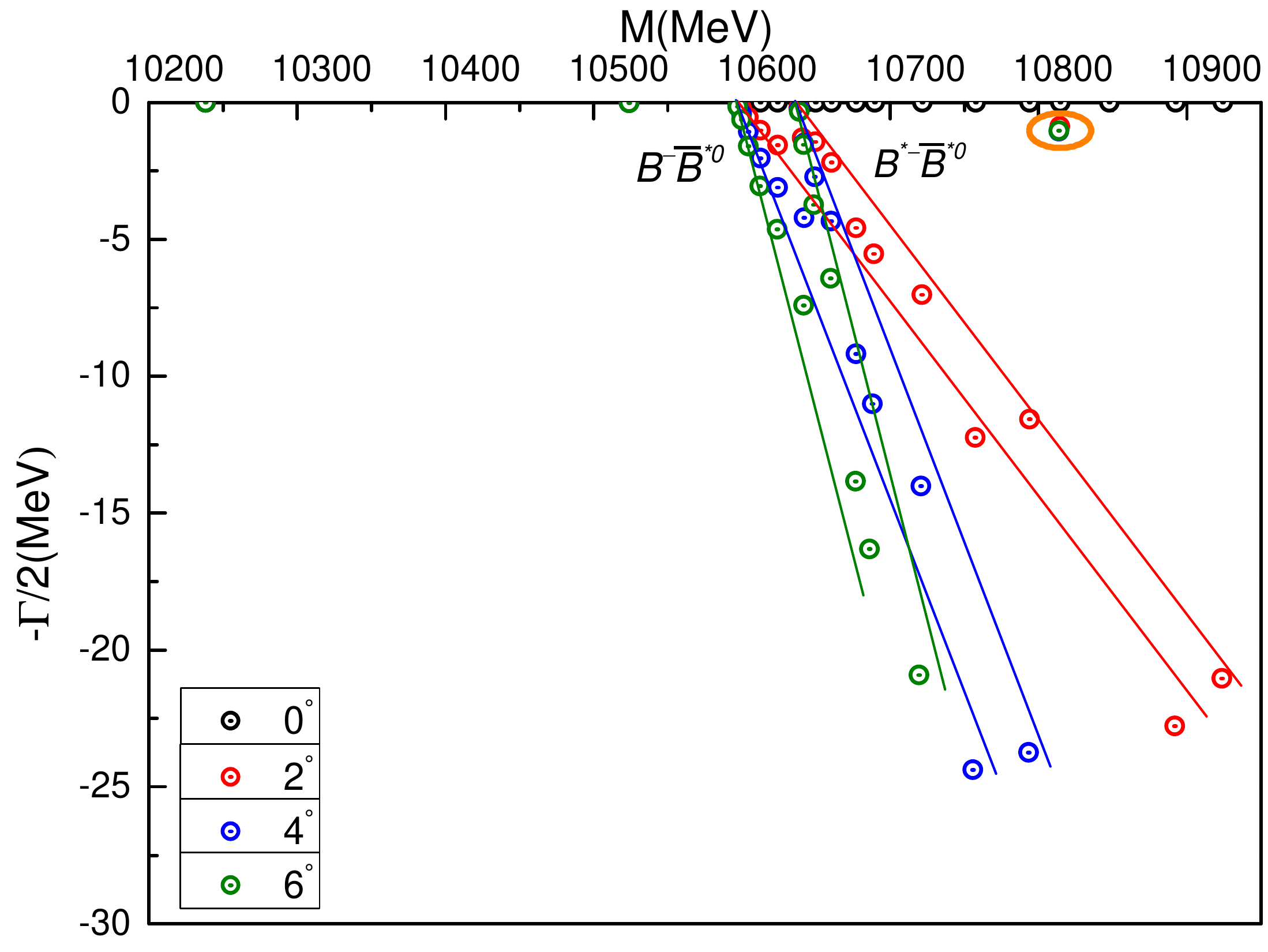}
\caption{Complex energies of doubly-bottom tetraquarks with $IJ^P=01^+$ in the coupled channels calculation, $\theta$ varying from $0^\circ$ to $6^\circ$ .} \label{PP2}
\end{figure}

The obtained deeply bound doubly-charm tetraquark with $M=3726$ MeV by CSM in the complete coupled channels calculation is clearly shown in Fig.~\ref{PP1}. We vary the rotated angle $\theta$ from $0^\circ$ to $6^\circ$, and this bound state remains on the real-axis. Particularly, the black dots in the real-axis are the calculated masses in coupled-channels calculation with $\theta=0^\circ$, and the red, blue and green ones are for complex energies with $\theta=2^\circ$, $4^\circ$ and $6^\circ$, respectively. Generally, they are aligned along the threshold lines with the same color and the nature of scattering state of $D^+D^{*0}$ and $D^{*+}D^{*0}$ in coupled-channels is clearly for their calculated poles always move along the cut lines when the scaling angle $\theta$ changes. However, there is a mismatch between the calculated dots and threshold lines in high energy region with large width. Nevertheless, we mainly focus on the low-lying state in this work and those calculation noises still present a nature of scattering states with obviously moving track.

In Fig.~\ref{PP1} one can see that there is a possible resonance pole marked with orange circle above the nearer $D^{*+}D^{*0}$ threshold lines. The three dots obtained by the CSM calculation with $\theta=2^\circ$, $4^\circ$ and $6^\circ$, respectively are located in a quite small energy region. Their complex energies are listed in Table~\ref{SumRest} and the estimated resonance mass and width is $\sim$4312 MeV and $\sim$16 MeV, respectively. By considering the fact that the resonance pole is near $D^{*+}D^{*0}$ threshold lines than $D^+D^{*0}$, hence the former channel should play a more important role in this resonance state.

\begin{table}[!t]
\caption{\label{GresultBB1} Lowest-lying states of doubly-bottom tetraquarks with quantum numbers $I(J^P)=0(1^+)$, unit in MeV.}
\begin{ruledtabular}
\begin{tabular}{lcccc}
Channel   & Color & $M$ & $E_B$ & $M'$ \\[2ex]
$B^- \bar{B}^{*0}$ & S   & $10585$ & $-12$  & $10592$ \\
$(10604)$           & H   & $10987$ & $+390$ & $10994$ \\
                    & S+H & $10562$ & $-35$  & $10569$ \\
                    & \multicolumn{4}{c}{Percentage (S;H): 83.0\%; 17.0\%} \\[2ex]
$B^{*-} \bar{B}^{*0}$ & S   & $10627$ & $-11$  & $10639$ \\
$(10650)$             & H   & $10974$ & $+336$ & $10986$ \\
                      & S+H & $10601$ & $-37$ & $10613$ \\
                      & \multicolumn{4}{c}{Percentage (S;H): 79.6\%; 20.4\%} \\[2ex] 
$(bb)^*(\bar{u}\bar{d})$ &    & $10261$ &  & \\[2ex] 
$(bb)(\bar{u}\bar{d})^*$ &    & $10787$ &  & \\[2ex]                   
Mixed  & & $10238^{1st}$ & & \\
                    & & $10524^{2nd}$ & & \\
\end{tabular}
\end{ruledtabular}
\end{table}

\begin{table}[!t]
\caption{{Component of each channel in coupled-channels calculation with $IJ^P=01^+$, the numbers $1$ and $8$ of superscript are for singlet-color and hidden-color channel respectively.}  \label{GresultCompBB}}
\begin{ruledtabular}
\begin{tabular}{lcccc}
  &  ~~$(B^- \bar{B}^{*0})^1$~~  & ~~$(B^{*-} \bar{B}^{*0})^1$~~   & ~~$(B^- \bar{B}^{*0})^8$~~ \\
 ~~$1st$~~  & ~~20.7\%~~  & ~~17.9\%~~  & 9.3\% \\
 ~~$2nd$~~  & ~~25.6\%~~  & ~~14.8\%~~  & 9.5\% \\[2ex]
  & ~~$(B^{*-} \bar{B}^{*0})^8$~~ & ~~$(bb)^*(\bar{u}\bar{d})$~~ & ~~$(bb)(\bar{u}\bar{d})^*$~~  \\ 
  ~~$1st$~~  & ~~9.4\%  & 42.6\%  & 0.1\% \\
  ~~$2nd$~~  & ~~9.1\% & 40.2\%  & 0.8\%  \\
\end{tabular}
\end{ruledtabular}
\end{table}

\begin{table}[!t]
\caption{\label{tab:disBB} The distance, in fm, between any two quarks of the found tetraquark bound-states in coupled-channels calculation, $(q=u,d)$.}
\begin{ruledtabular}
\begin{tabular}{cccc}
 & $r_{\bar{u}\bar{d}}$ & $r_{\bar{q}b}$ & $r_{bb}$  \\[2ex]
 $1st$ & 0.604 & 0.608 & 0.328 \\
 $2nd$ & 0.830 & 0.734 & 0.711 \\
\end{tabular}
\end{ruledtabular}
\end{table}

\subsection{doubly-bottom tetraquarks}
We herein investigate $B^{(*)-} \bar{B}^{*0}$ and $(bb)^{(*)}(\bar{u}\bar{d})^{(*)}$ channels which are similar to the doubly-charm tetraquarks. Possible bound and resonance states are also obtained only in $I(J^P)=0(1^+)$ state. However, with much more heavier $b$-flavored quarks included, possible bound states of color-singlet channels of $B^- \bar{B}^{*0}$ and $B^{*-} \bar{B}^{*0}$ are found, their binding energies are $-12\,\text{MeV}$ and $-11\,\text{MeV}$ respectively. Additionally, in Table~\ref{GresultBB1} one can find that nearly triple binding energies are obtained both for $B^- \bar{B}^{*0}$ ($E_B=-35\,\text{MeV}$) and $B^{*-} \bar{B}^{*0}$ ($E_B=-37\,\text{MeV}$) when the hidden-color channels are incorporated in the calculation. These deeper bound states than $D^{(*)+} D^{*0}$ cases also indicate a strong coupling which is about 80\% color-singlet component for $B^{(*)-} \bar{B}^{*0}$. After a mass shift for these two bound states, the slightly modified masses of doubly-bottom tetraquarks are $10569\,\text{MeV}$ and $10613\,\text{MeV}$ respectively.

In diquark-antidiquark configuration, according to the $B^- \bar{B}^{*0}$ theoretical thresholds, one tightly bound state of $(bb)^*(\bar{u}\bar{d})$ whose binding energy is $E_B=-336\,\text{MeV}$ and one excited state of $(bb)(\bar{u}\bar{d})^*$ with $E_B=+190\,\text{MeV}$ are shown in Table~\ref{GresultBB1}, respectively. This situation is also consistent with $(cc)^{(*)}(\bar{u}\bar{d})^{(*)}$ channels which are of smaller binding energies. The obtained deeply bound state $(bb)^*(\bar{u}\bar{d})$ at $10261\,\text{MeV}$ is supported by Refs.~\cite{ejecq:2017prl, mkjlr:2017prl, cefgk:2019prd, jclh:1988prd}, only $\sim$$130\,\text{MeV}$ lower than the predicted mass in Ref.~\cite{mkjlr:2017prl}.

Furthermore, two bound states are found in a coupled-channels calculation which all the channels listed in Table~\ref{GresultBB1} are considered, their masses are $10238\,\text{MeV}$ and $10524\,\text{MeV}$, respectively. Clearly, the $(bb)^*(\bar{u}\bar{d})$ diquark-antidiquark channel is pushed down by $23\,\text{MeV}$ due to the coupling effect, and the second bound state ($M=10524\,\text{MeV}$) is $73\,\text{MeV}$ below the $B^- \bar{B}^{*0}$ theoretical threshold. Then with a purpose of disentangling the nature of these two obtained bound states, their components and inner structures are studied. One can see in Table~\ref{GresultCompBB} that the components of the two bound states are quite comparable and both about 42\% for $(bb)^*(\bar{u}\bar{d})$ channel and about 20\% sub-dominant for the color-singlet channel of $B^{(*)-} \bar{B}^{*0}$. With no more than $0.83\,\text{fm}$ distance for any quark pair listed in Table~\ref{tab:disBB}, the compact tetraquark structures for these two bound states are clearly presented again, and one need to mention that the distances of two bottom quarks for them are only $0.328\,\text{fm}$ and $0.711\,\text{fm}$, respectively.

In Table~\ref{SumRest} and Fig.~\ref{PP2} one can find that the two bound states are stable against the change of scaling angle $\theta$. Besides, one resonance state which mass and width is $10814\,\text{MeV}$ and $2\,\text{MeV}$, respectively is obtained in the complete coupled-channels calculation with various rotated angle $\theta$. We mark it with a big orange circle where the three dots are almost overlap and their complex energies within $\theta$ taken the value of $2^\circ$, $4^\circ$ and $6^\circ$ are listed in Table~\ref{SumRest}, respectively. This narrow width resonance pole is close to $B^{*-} \bar{B}^{*0}$ threshold line and more contributions should be made by this channel. However, the other poles with a scattering nature are generally aligned along the $B^- \bar{B}^{*0}$ and $B^{*-} \bar{B}^{*0}$ threshold lines.

%%%%%%%%%%
\begin{table}[!t]
\caption{\label{GresultCB1} Lowest-lying states of charm-bottom tetraquarks with quantum numbers $I(J^P)=0(0^+)$, unit in MeV.}
\begin{ruledtabular}
\begin{tabular}{lcccc}
Channel   & Color & $M$ & $E_B$ & $M'$ \\[2ex]
$D^0 \bar{B}^0$ & S   & $7172$ & $-4$  & $7143$ \\
$(7147)$         & H   & $7685$ & $+509$ & $7656$ \\
                  & S+H & $7171$ & $-5$  & $7142$ \\
                  & \multicolumn{4}{c}{Percentage (S;H): 96.4\%; 3.6\%}  \\[2ex]
$D^{*0} \bar{B}^{*0}$ & S   & $7327$ & $-9$  & $7325$ \\
$(7334)$           & H   & $7586$ & $+250$ & $7584$ \\
                    & S+H & $7297$ & $-39$  & $7295$ \\
                    & \multicolumn{4}{c}{Percentage (S;H): 87.8\%; 12.2\%} \\[2ex]
$(cb)(\bar{u}\bar{d})$ &    & $7028$ &  & \\[2ex]
$(cb)^*(\bar{u}\bar{d})^*$ &    & $7482$ &  & \\[2ex]
Mixed  & & $6980$ & & \\
\end{tabular}
\end{ruledtabular}
\end{table}

\begin{figure}[ht]
\epsfxsize=3.5in \epsfbox{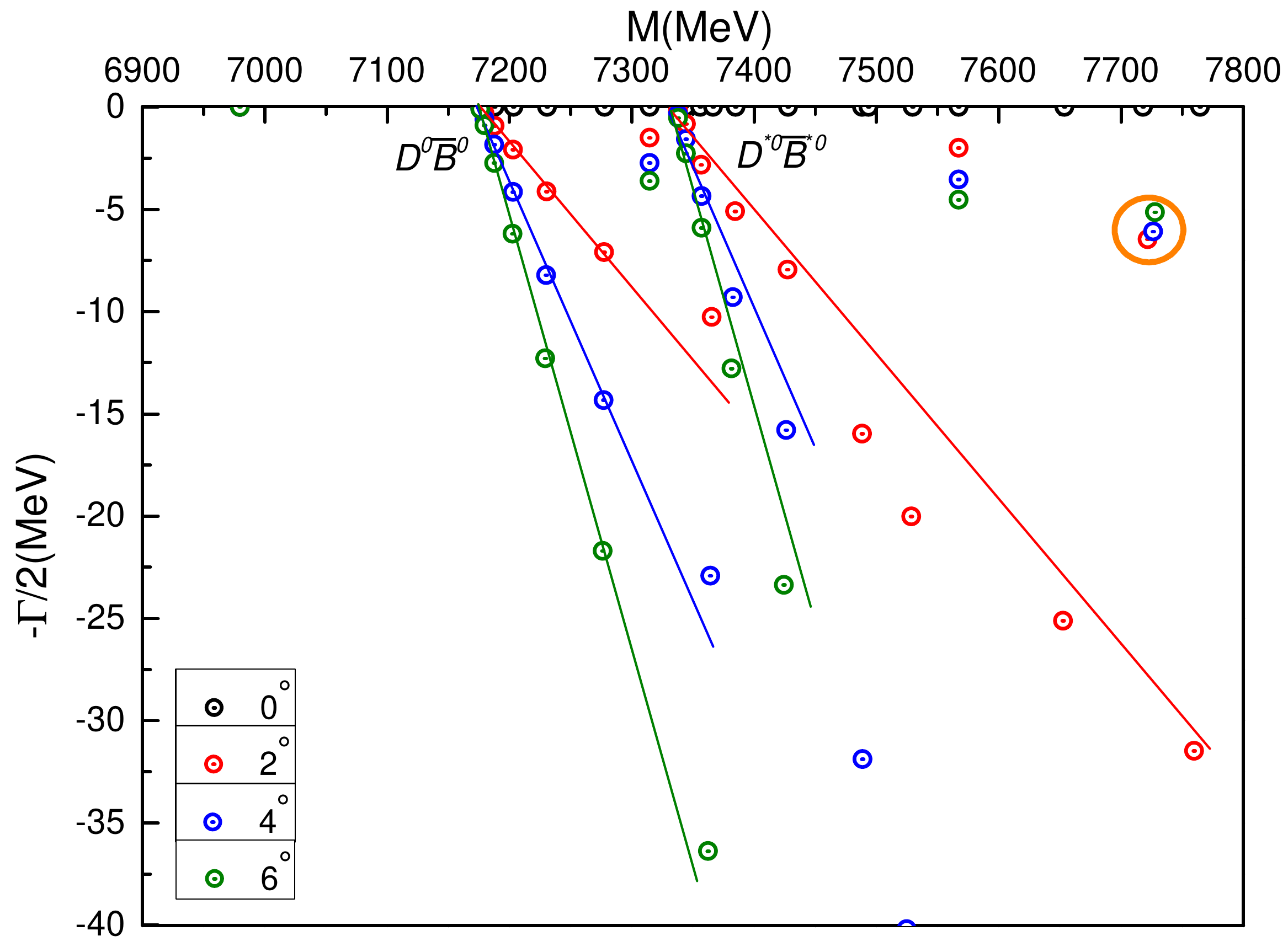}
\caption{Complex energies of charm-bottom tetraquarks with $IJ^P=00^+$ in the coupled channels calculation, $\theta$ varying from $0^\circ$ to $6^\circ$ .} \label{PP3}
\end{figure}

\subsection{charm-bottom tetraquarks}
In these sector, some bound or resonance states are obtained only for iso-scalar tetraquarks, and our theoretical findings in meson-meson channels are comparable with those results in Table V of Ref.~\cite{tfcjvav2019}. Hence we will discuss them according to $I(J^P)$ quantum numbers individually.

{\bf The $\bm{I(J^P)=0(0^+)}$ channel:}
Loosely bound states of the color-singlet channel of $D^0 \bar{B}^0$ and $D^{*0} \bar{B}^{*0}$ are found, their binding energies are $-4\,\text{MeV}$ and $-9\,\text{MeV}$, respectively. In Table~\ref{GresultCB1} one can realize that there is only a remarkable coupling effect ($E_B=-39\,\text{MeV}$) on $D^{*0} \bar{B}^{*0}$ configuration when the hidden-color channel is incorporated, and almost no influence on $D^0 \bar{B}^0$ channel with only $1\,\text{MeV}$ binding energy increased. This is supported by our calculated proportion for color-singlet and hidden-color channels: 96.4\% for $(D^0 \bar{B}^0)^1$ and 87.8\% for $(D^{*0} \bar{B}^{*0})^1$. Meanwhile, one deeply bound state $(cb)(\bar{u}\bar{d})$ with $E_B=-148\,\text{MeV}$ and one excited state $(cb)^*(\bar{u}\bar{d})^*$ with $E_B=+306\,\text{MeV}$ are found with respect to the $D^0 \bar{B}^0$ theoretical threshold. The binding energy of lowest-lying state is increased by $48\,\text{MeV}$ in the complete coupled-channels calculation. This tightly bound state which mass is $6980\,\text{MeV}$ brings us a compact doubly-heavy tetraquark structure again, Table~\ref{tab:disDB1} presents the size of state around $0.6\,\text{fm}$ and even smaller distance, $0.428\,\text{fm}$ for $cb$ quark pair.  All of these features can be related to the strong coupling effect which almost 50\% for $(cb)(\bar{u}\bar{d})$, 26.4\% for $(D^0 \bar{B}^0)^1$ and 21.5\% for $(D^{*0} \bar{B}^{*0})^1$ channels are shown in Table~\ref{GresultCompDB1}.

In the complex scaling computation that the investigated region of rotated angle $\theta$ is the same as previous two types of tetraquark states, the bound state along with a resonance are presented in Fig.~\ref{PP3}. Specifically, four dots whose $\theta$ taken the value of $0^\circ$, $2^\circ$, $4^\circ$ and $6^\circ$, respectively are overlap exactly at mass is $6980\,\text{MeV}$ on the real-axis. The resonance pole is found near the mass of $7726\,\text{MeV}$ and its width is $\sim$$12\,\text{MeV}$ according to Table~\ref{SumRest}. Moreover, one can find in Fig.~\ref{PP3} that the resonance state is far from the $D^0 \bar{B}^0$ threshold and accordingly, the majority contributions should owing to $D^{*0} \bar{B}^{*0}$ channel.

The rest calculated poles in Fig.~\ref{PP3} are basically fit well with the $D^0 \bar{B}^0$ and $D^{*0} \bar{B}^{*0}$ threshold lines, except for two cases. Namely, the dots always descend slowly with the increasing of scaling angle $\theta$ both at mass is $7314\,\text{MeV}$ and $7567\,\text{MeV}$. They can not be identified as resonance states due to the instability.

\begin{table}[!t]
\caption{\label{GresultCB2} Lowest-lying states ofcharm-bottom tetraquarks with quantum numbers $I(J^P)=0(1^+)$, unit in MeV.}
\begin{ruledtabular}
\begin{tabular}{lcccc}
Channel   & Color & $M$ & $E_B$ & $M'$ \\[2ex]
$D^0 \bar{B}^{*0}$ & S   & $7214$ & $-3$  & $7190$ \\
$(7193)$           & H   & $7694$ & $+477$ & $7670$ \\
                    & S+H & $7213$ & $-4$  & $7189$ \\
                    & \multicolumn{4}{c}{Percentage (S;H): 96.8\%; 3.2\%} \\[2ex]
$D^{*0} \bar{B}^0$ & S   & $7293$ & $-2$ & $7286$ \\
$(7288)$           & H   & $7707$ & $+412$ & $7700$ \\
                    & S+H & $7292$ & $-3$ & $7285$ \\
                    & \multicolumn{4}{c}{Percentage (S;H): 96.8\%; 3.2\%} \\[2ex]
$D^{*0} \bar{B}^{*0}$ & S   & $7334$ & $-2$  & $7332$ \\
$(7334)$             & H   & $7691$ & $+354$ & $7688$ \\
                      & S+H & $7326$ & $-10$ & $7324$ \\
                      & \multicolumn{4}{c}{Percentage (S;H): 89.3\%; 10.7\%} \\[2ex] 
$(cb)^*(\bar{u}\bar{d})$ &    & $7039$ &  & \\[2ex] 
$(cb)(\bar{u}\bar{d})^*$ &    & $7531$ &  & \\[2ex] 
$(cb)^*(\bar{u}\bar{d})^*$ &    & $7507$ &  & \\[2ex]                   
Mixed  & & $6997$ & & \\
\end{tabular}
\end{ruledtabular}
\end{table}

\begin{figure}[ht]
\epsfxsize=3.5in \epsfbox{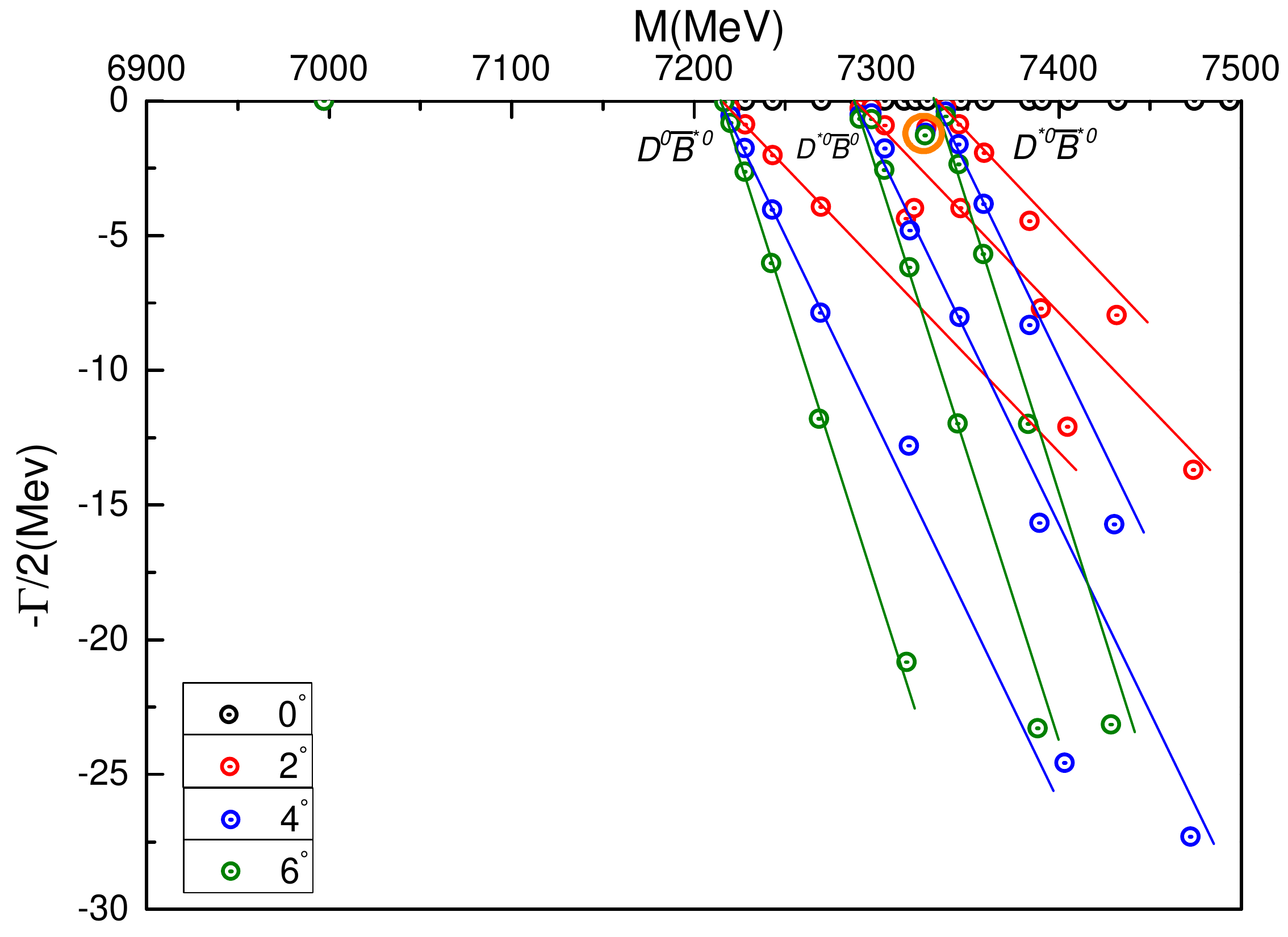}
\caption{Complex energies of charm-bottom tetraquarks with $IJ^P=01^+$ in the coupled channels calculation, $\theta$ varying from $0^\circ$ to $6^\circ$.} \label{PP4}
\end{figure}

{\bf The $\bm{I(J^P)=0(1^+)}$ channel:}
Both of three channels in meson-meson $D^{(*)0} \bar{B}^{(*)0}$ and diquark-antidiquark $(cb)^{(*)}(\bar{u}\bar{d})^{(*)}$ configurations are studied in Table~\ref{GresultCB2}. Four similar features as the other doubly-heavy tetraquarks discussed before can be drawn: (i) loosely bound states with $E_B=-3\,\text{MeV}$, $-2\,\text{MeV}$ and $-2\,\text{MeV}$ for the three color-singlet channels of $D^0 \bar{B}^{*0}$, $D^{*0} \bar{B}^0$ and $D^{*0} \bar{B}^{*0}$ respectively, (ii) the coupling between color-singlet and hidden-color channels are quite weak ($E_B$ increased by $1\,\text{MeV}$) for $D^0 \bar{B}^{*0}$ and $D^{*0} \bar{B}^0$ configurations, but $8\,\text{MeV}$ increased binding energy for $D^{*0} \bar{B}^{*0}$, (iii) only one deeply bound state in single channel calculation, namely $E_B=-178\,\text{MeV}$ for $(cb)^*(\bar{u}\bar{d})$ channel when compared with the lowest theoretical threshold of $D^0 \bar{B}^{*0}$, and (iV) more tightly bound state which mass is $6997\,\text{MeV}$ in the complete coupled-channels calculation.

In Table~\ref{GresultCompDB1} one can see that the most contribution 46.4\% comes from $(cb)^*(\bar{u}\bar{d})$ channel and other three sub-dominant channels are 20.2\% for $(D^0 \bar{B}^{*0})^1$, 11.6\% for $(D^{*0} \bar{B}^0)^1$ and 16.8\% for $(D^{*0} \bar{B}^{*0})^1$. These facts of strong coupling effect along with the domination of diquark-antidiquark configuration result in a compact structure again, and one can find a comparable size between $IJ^P=01^+$ and $00^+$ state in Table~\ref{tab:disDB1}.

Fig.~\ref{PP4} presents the distribution of complex energies in the complete coupled-channels calculation. Three scattering states of $D^0 \bar{B}^{*0}$, $D^{*0} \bar{B}^0$ and $D^{*0} \bar{B}^{*0}$ are clearly shown and the bound state which mass is $6997\,\text{MeV}$ remains on the real-axis. Meanwhile, one narrow width resonance state as doubly-bottom tetraquarks whose $\Gamma=2\,\text{MeV}$ is obtained and marked with a orange circle in the figure. $D^{*0} \bar{B}^0$ and $D^{*0} \bar{B}^{*0}$ channels should be both important to this quite narrow resonance pole which is among the threshold lines of them. The resonance mass is $7327\,\text{MeV}$ and its width is $\sim$$2.4\,\text{MeV}$ in the CSM computation with $\theta$ varying from $0^\circ$ to $6^\circ$ in Table~\ref{SumRest}.  

\begin{table}[!t]
\caption{\label{GresultCB3} Lowest-lying states of charm-bottom tetraquarks with quantum numbers $I(J^P)=0(2^+)$, unit in MeV.}
\begin{ruledtabular}
\begin{tabular}{lcccc}
Channel   & Color & $M$ & $E_B$ & $M'$ \\[2ex]
$D^{*0} \bar{B}^{*0}$ & S   & $7334$ & $-2$  & $7332$ \\
$(7334)$             & H   & $7720$ & $+384$ & $7718$ \\
                      & S+H & $7334$ & $-2$  & $7332$ \\
                      & \multicolumn{4}{c}{Percentage (S;H): 99.8\%; 0.2\%} \\[2ex]
$(cb)^*(\bar{u}\bar{d})^*$ &    & $7552$ &  & \\[2ex]                   
Mixed  & & $7333$ & & \\
\end{tabular}
\end{ruledtabular}
\end{table}

\begin{figure}[ht]
\epsfxsize=3.5in \epsfbox{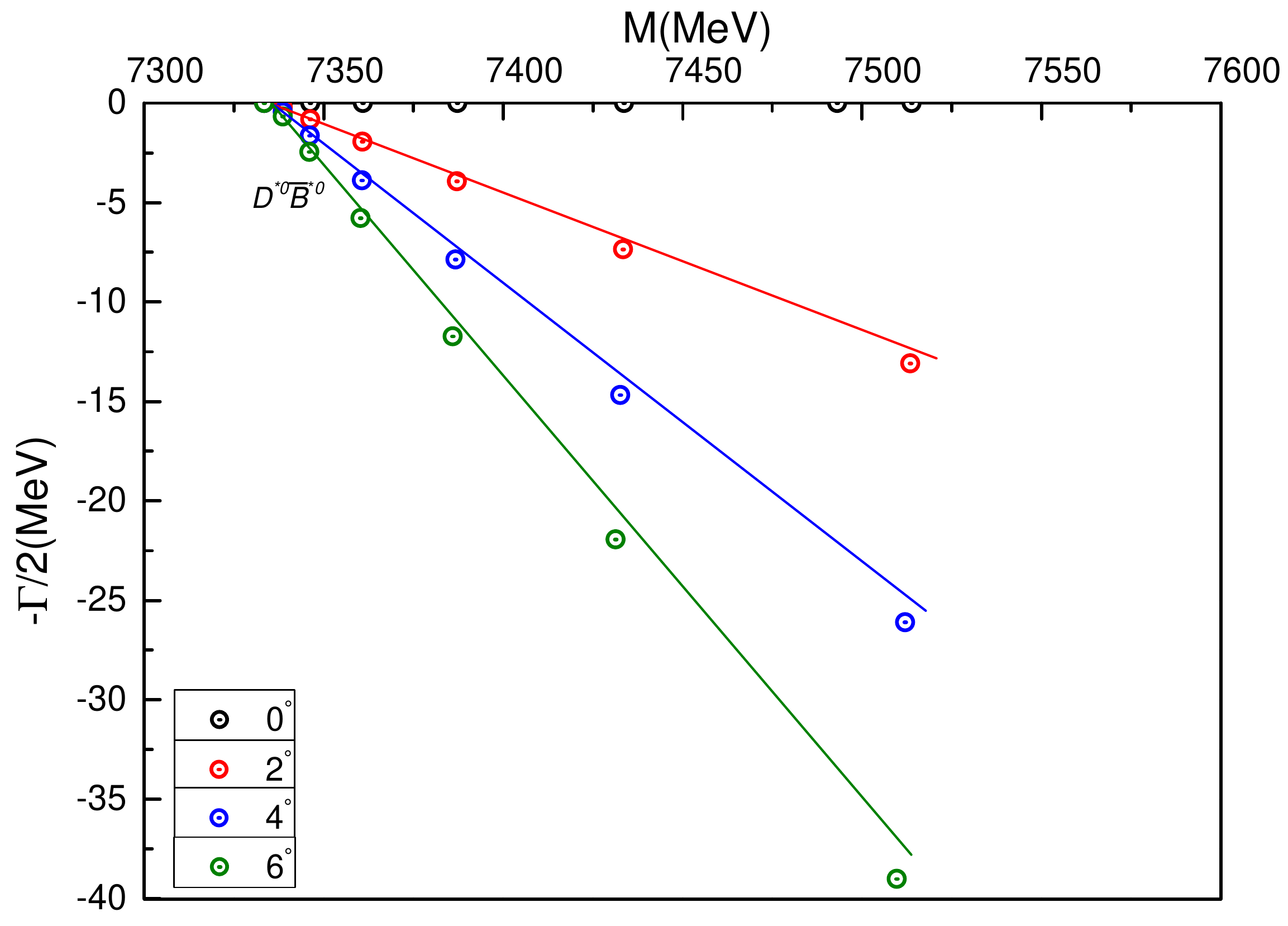}
\caption{Complex energies of charm-bottom tetraquarks with $IJ^P=02^+$ in the coupled channels calculation, $\theta$ varying from $0^\circ$ to $6^\circ$ .} \label{PP5}
\end{figure}

{\bf The $\bm{I(J^P)=0(2^+)}$ channel:}
Only two channels contribute to this case: $D^{*0} \bar{B}^{*0}$ meson-meson channel and diquark-antidiquark one $(cb)^*(\bar{u}\bar{d})^*$. As in all cases studied before, a loosely bound state of color-singlet channel $D^{*0} \bar{B}^{*0}$ is obtained with $E_B=-2\,\text{MeV}$. Furthermore, the coupling is still quite weak in the complete coupled-channels investigation for $(D^{*0} \bar{B}^{*0})^1$ channel contributes 98.6\%, and the calculated mass is $7333\,\text{MeV}$ which is quite close to the color-singlet channel one of $7334\,\text{MeV}$. This indicates the nature of molecular-type meson-meson structure and it is also consistent with the obtained size in Table~\ref{tab:disDB1} 
where the distances between any two quarks are about $1.6\,\text{fm}$ $\sim$ $2.2\,\text{fm}$.

In additional, no resonance state is found in the complete coupled-channels calculation with $\theta$ varying from $0^\circ$ to $6^\circ$. The loosely bound state with $M=7333\,\text{MeV}$ and another scattering state of $D^{*0} \bar{B}^{*0}$ are presented in Fig.~\ref{PP5}, respectively.

\begin{table}[!t]
\caption{{Component of each channel in coupled-channels calculation, the numbers $1$ and $8$ of superscript are for singlet-color and hidden-color channel respectively, $(q=u,d)$.}  \label{GresultCompDB1}}
\begin{ruledtabular}
\begin{tabular}{lcccc}
 $IJ^P$ & ~~$(D^0 \bar{B}^0)^1$~~  & ~~$(D^{*0} \bar{B}^{*0})^1$~~   & ~~$(D^0 \bar{B}^0)^8$~~ &
   ~~$(D^{*0} \bar{B}^{*0})^8$~~ \\
$00^+$ & ~~26.4\%~~  & ~~21.5\%~~  & 1.6\%~~  & 1.9\% ~~\\[2ex]
 & ~~$(cb)(\bar{u}\bar{d})$~~ & ~~$(cb)^*(\bar{u}\bar{d})^*$~~  \\ 
 & ~~48.5\%~~  & ~~0.1\%~~ \\[2ex]
 $01^+$ & ~~$(D^0 \bar{B}^{*0})^1$~~  & ~~$(D^{*0} \bar{B}^0)^1$~~   & ~~$(D^{*0} \bar{B}^{*0})^1$~~ &
   ~~$(D^0 \bar{B}^{*0})^8$~~ \\
 & ~~20.2\%~~  & ~~11.6\%~~  & ~~16.8\%~~  & ~~1.4\% ~~\\[2ex]
 & ~~$(D^{*0} \bar{B}^0)^8$~~ & ~~$(D^{*0} \bar{B}^{*0})^8$~~ & ~~$(cb)^*(\bar{u}\bar{d})$~~ &
  ~~$(cb)(\bar{u}\bar{d})^*$~~  \\ 
 & ~~1.3\%~~  & 1.8\%~~ & ~~46.4\%~~  & ~~0.1\%~~ \\[2ex]
 & ~~$(cb)^*(\bar{u}\bar{d})^*$~~ \\
 & ~~0.4\%~~ \\[2ex]
 $02^+$ & ~~$(D^{*0} \bar{B}^{*0})^1$~~  & ~~$(D^{*0} \bar{B}^{*0})^8$~~   & ~~$(cb)^*(\bar{u}\bar{d})^*$~~\\
 & ~~98.6\%~~  & ~~0.3\%~~  & ~~1.1\%~~\\
\end{tabular}
\end{ruledtabular}
\end{table}

\begin{table}[!t]
\caption{\label{tab:disDB1} The distance, in fm, between any two quarks of the found tetraquark bound-states in coupled-channels calculation, $(q=u,d)$.}
\begin{ruledtabular}
\begin{tabular}{ccccc}
  $IJ^P$& $r_{\bar{u}\bar{d}}$ & $r_{\bar{q}c}$ & $r_{\bar{q}b}$  & $r_{cb}$  \\[2ex]
  $00^+$ & 0.635 & 0.653 & 0.610  & 0.428 \\[2ex]
  $01^+$ & 0.632 & 0.661 & 0.616  & 0.434 \\[2ex]
  $02^+$ & 2.248 & 1.612 & 1.597  & 2.102 \\
\end{tabular}
\end{ruledtabular}
\end{table}

\begin{table*}[!t]
\caption{\label{SumRest} Possible bound and resonance states for $QQ\bar{q}\bar{q}$ $(q=u~or~d)$ tetraquarks in CSM with rotated angle $\theta$ varying from $0^\circ$ to $6^\circ$. The imaginary part of complex energy and resonance width are with the relation of Im$(E)=-\Gamma/2$, unit in MeV.}
\begin{ruledtabular}
\begin{tabular}{cccccc}
   & & $0^\circ$ & $2^\circ$ & $4^\circ$ & $6^\circ$~~ \\
$cc\bar{q}\bar{q}$~~&~bound state~ & 3726 & 3726 & 3726 & 3726~~ \\
$IJ^P=01^+$~~&~resonance state~ & - & $4319-7.9i$ & $4312-9.4i$ & $4310-7.3i$~~ \\[2ex]
$bb\bar{q}\bar{q}$~~&~bound state~ & 10238; 10524 & 10238; 10524 & 10238; 10524 & 10238; 10524~~ \\
$IJ^P=01^+$~~&~resonance state~ & - & $10814-0.9i$ & $10814-1.1i$ & $10814-1.0i$~~ \\[2ex]
$cb\bar{q}\bar{q}$~~&~bound state~ & 6980 & 6980 & 6980 & 6980~~ \\
$IJ^P=00^+$~~&~resonance state~ & - & $7722-6.5i$ & $7726-6.1i$ & $7728-5.2i$~~ \\[2ex]
$cb\bar{q}\bar{q}$~~&~bound state~ & 6997 & 6997 & 6997 & 6997~~ \\
$IJ^P=01^+$~~&~resonance state~ & - & $7327-1.0i$ & $7327-1.2i$ & $7327-1.3i$~~ \\[2ex]
$cb\bar{q}\bar{q}$~~&~bound state~ & 7333 & 7333 & 7333 & 7333~~ \\
$IJ^P=02^+$~~&~resonance state~ & - &-  &-  &- ~~ 
\end{tabular}
\end{ruledtabular}
\end{table*}

%%%%%%%%%%%%%%%%%%%%%%%%%%%%%%%%%%%%%%%%%%%%%%%%%%%%%%%%%%%%%%%%%%%%%%%%%%%%%%%%%%%%%%%%%%

\section{Epilogue}
\label{sec:summary}

%With two dozens of exotic XYZ mesons observed at different high energy collision facilities, extensive theoretical investigations have been devoted to the four-quark sectors. There are many debates on the existence of bound states in fully-heavy tetraquarks, however, results on doubly-heavy tetraquarks are more compatible. A compact $bb\bar{u}\bar{d}$ state with $IJ^P=01^+$ is predicted~\cite{ejecq:2017prl, mkjlr:2017prl, cefgk:2019prd, jclh:1988prd}, and axial-vector tetraquark state $bc\bar{u}\bar{d}$ at mass $7105\pm 155\,\text{MeV}$~\cite{ssaka:2019arx}.

In a complex scaling range of chiral quark formalism, by considering meson-meson and diquark-antidiquark configurations along with all color structures (couplings are also considered), $i. e.$ color-singlet and hidden-color channels for dimeson structure; color triplet-antitriplet and sextet-antisextet channels for $(QQ)(\bar{q}\bar{q})$ structure, we have studied the possibility of having tetraquark bound- and resonance-states in the doubly-heavy sectors with quantum numbers $J^P=0^+$, $1^+$ and $2^+$, and in the $0$ and $1$ isospin sectors. For possible bound states in the complete coupled-channels study, their inner structures and components are also analyzed by computing the distances among any pair of quarks and the contributions of each channel's wave functions. Masses and widths for possible resonance states are also calculated in the coupled-channels calculation. The model parameters which are included in the perturbative one-gluon exchange, the nonperturbative linear-screened confining and Goldstone-boson exchange interactions between light quarks have been fitted in the past through hadron, hadron-hadron and multiquark phenomenology.

For all quantum states of the investigated doubly-heavy tetraquarks, $cc\bar{q}\bar{q}$, $bb\bar{q}\bar{q}$ and $cb\bar{q}\bar{q}$ $(q=u, d)$, tightly bound and narrow resonance states are only obtained in $IJ^P=01^+$ state for the former two sectors, and they are also obtained for $cb\bar{q}\bar{q}$ in $00^+$ and $01^+$ states. However, only loosely bound state is found for charm-bottom tetraquarks in $02^+$ states. All of these states within meson-meson configurations are loosely bound whether in color-singlet channels or coupling to hidden-color ones. However, compact structures are available in diquark-antidiquark channels except for charm-bottom tetraquarks in $02^+$ states. Let us characterize the features in detail.

 Firstly, in doubly-charm tetraquark states, two loosely bound states $D^+D^{*0}$ and $D^{*+}D^{*0}$ with mass $3876\,\text{MeV}$ and $4017\,\text{MeV}$, respectively are obtained in $IJ^P=01^+$ state. Meanwhile, a deeply bound state with $(cc)^*(\bar{u}\bar{d})$ diquark-antidiquark structure is found at  $3778\,\text{MeV}$. In the complete coupled-channels calculation the lowest-lying state mass is $3726\,\text{MeV}$, and the compact tetraquark states size is $0.52-0.66\,\text{fm}$. Meanwhile, a resonance state which is mainly induced by $D^{*+}D^{*0}$ channel is obtained and the estimated mass and width is $4312\,\text{MeV}$ and $16\,\text{MeV}$, respectively.

 Secondly, similar to the doubly-charm tetraquarks, we found loosely bound states of $B^-\bar{B}^{*0}$ and $B^{*-}\bar{B}^{*0}$ with $IJ^P=01^+$, the predicted masses are $10569\,\text{MeV}$ and $10613\,\text{MeV}$, respectively. There are $\sim$20\% contributions from hidden-color channels for these two molecular states. Diquark-antidiquark state $(bb)^*(\bar{u}\bar{d})$ is much more tightly bound with a binding energy $E_B=-336\,\text{MeV}$ when compares with the theoretical threshold of $B^-\bar{B}^{*0}$ channel. In the complete coupled-channels calculation, two compact tetraquark bound states with mass at $10238\,\text{MeV}$ and $10524\,\text{MeV}$, respectively are obtained. The distances among any quark pair of them are less than $0.83\,\text{fm}$. Besides, a narrow resonance state with mass $M=10814\,\text{MeV}$ and width $\Gamma=2\,\text{MeV}$ is found, and $B^{*-}\bar{B}^{*0}$ channel plays an important role to this state.

 In additional, possible charm-bottom tetraquark states are found in three quantum states $IJ^P=00^+$, $01^+$ and $02^+$. Specifically, in $00^+$ state, $D^0 \bar{B}^0(7142)$ and $D^{*0} \bar{B}^{*0}(7295)$; in $01^+$ state, $D^0 \bar{B}^{*0}(7189)$, $D^{*0} \bar{B}^0(7285)$ and $D^{*0} \bar{B}^{*0}(7324)$; and in $02^+$ state, $D^{*0} \bar{B}^{*0}(7332)$, the predicted masses for these molecular states are correspondingly signed in the brackets. The compact tetraquarks $(cb)(\bar{u}\bar{d})$ and $(cb)^*(\bar{u}\bar{d})$ with mass at $7028\,\text{MeV}$ and $7039\,\text{MeV}$ are found in $00^+$ and $01^+$ states, respectively. In the complete coupled-channels calculation, these two states are of lower masses $6980\,\text{MeV}$ and $6997\,\text{MeV}$, besides their size are both less than $0.67\,\text{fm}$. However, $02^+$ state remains the molecular type structure due to quite weak coupling. Two resonances are available for $00^+$ and $01^+$ states, their mass and width are $7726\,\text{MeV}$, $12\,\text{MeV}$ and $7327\,\text{MeV}$, $2.4\,\text{MeV}$, respectively. $D^{*0} \bar{B}^{*0}$ channel is crucial for the resonance state with $IJ^P=00^+$ and resonance in $01^+$ state is mainly induced by $D^{*0} \bar{B}^{0}$ and $D^{*0} \bar{B}^{*0}$ channels.

Finally, our results in this work by the phenomenological framework of chiral quark model are expecting to be confirmed in future high energy experiments. Meanwhile, a natural extension of our investigation in next step will be the other open-heavy tetraquark states, $i. e.$ $QQ\bar{Q}\bar{q}$ systems. Properties in those almost non-relativistic systems are also absorbing.
%by means of complex scaling method which has been extensively applied to nuclear physics problems~\cite{SAPTP11612006, TMPPNP7912014}. Not only the bound state, but also the resonance one can be studied within one scheme. The application of this technique in tetraquark sectors can be expected in our following study.

%%%%%%%%%%%%%%%%%%%%%%%%%%%%%%%%%%%%%%%%%%%%%%%%%%%%%%%%%%%%%%%%%%%%%%%%%%%%%%%%%%%%%%%%%%

\begin{acknowledgments}
G. Yang would like to thank L. He for his support and informative discussions.
Work partially financed by: China Postdoctoral Science Foundation Grant no. 2019M650617; National Natural Science Foundation of China under Grant nos. 11535005 and 11775118; 
%European Union's Horizon 2020 research and innovation programme under the Marie Sk\l{}odowska-Curie grant agreement no. 665919; Spanish MINECO's Juan de la Cierva-Incorporaci\'on programme with grant agreement no. IJCI-2016-30028; and by Spanish Ministerio de Econom\'ia, Industria y Competitividad under contract nos. FPA2014-55613-P, FPA2017-86989-P and SEV-2016-0588.
Spanish Ministerio de Econom\'ia, Industria y Competitividad under contract no. FPA2017-86380-P.
\end{acknowledgments}

%%%%%%%%%%%%%%%%%%%%%%%%%%%%%%%%%%%%%%%%%%%%%%%%%%%%%%%%%%%%%%%%%%%%%%%%%%%%%%%%%%%%%%%%%%

%\bibliography{Doublyheavytetraquarks}

\end{document}